\documentclass[12pt,a4paper,journal,onecolumn]{IEEEtran}
\pdfoutput=1
\usepackage{cite}

\ifCLASSINFOpdf
  \usepackage[pdftex]{graphicx}
  \graphicspath{{../pdf/}{../jpeg/}}
  \DeclareGraphicsExtensions{.pdf,.jpeg,.png}
\else
  \usepackage[dvips]{graphicx}
\fi

\usepackage[cmex10]{amsmath}
\usepackage{amssymb}

\usepackage{amsthm}

\usepackage[font=footnotesize]{subfig}

\newcommand{\tfa}{time-frequency analysis}

\newcommand{\tf}{time-frequency}

\newcommand{\beqa}{\begin{eqnarray*}}
\newcommand{\eeqa}{\end{eqnarray*}}

\newcommand{\field}[1]{\mathbb{#1}}
\newcommand{\bR}{\field{R}}        
\newcommand{\bN}{\field{N}}        
\newcommand{\bZ}{\field{Z}}        
\newcommand{\bC}{\field{C}}        
\newcommand{\bT}{\field{T}}        %
 \def\cH{\mathcal{H}}

 \def\cG{\mathcal{G}}

 \def\cT{\mathcal{T}}

\def\hg{\hat{g}}

\def\ltZ{\ell^2(\bZ )}
\def\Lt{L^2}
\def\Lo{L^1}
\def\lt{\ell^2}

\def\<{\left<}
\def\>{\right>}

\def\absl{\bigl\lvert}
\def\absr{\bigr\rvert}

\def\inv{^{-1}}

\def\mv1{M_v^1}

\newtheorem{theorem}{Theorem}
\newtheorem{proposition}[theorem]{Proposition}
\newtheorem{lemma}[theorem]{Lemma}
\newtheorem{corollary}[theorem]{Corollary}

\newtheorem{definition}[theorem]{Definition}

\newtheorem{remark}[theorem]{Remark}
\newtheorem{example}[theorem]{Example}
\newtheorem{algorithm}[theorem]{Algorithm}
\newtheorem{claim}[theorem]{Claim}

\newcommand{\gab}{\cG (g,\alpha ,\beta )}

\begin{document}
%
\title{Discretized Gabor Frames of Totally Positive Functions}
%
%
%

\author{Severin Bannert, Karlheinz Gr\"ochenig, and Joachim St\"ockler
\thanks{S.\ B.\ was supported by WWTF grant ICT 10-066}
\thanks{K.\ G.\ acknowledges the 
  support of the   projects P22746-N13 and  SISE -- S10602  of the
Austrian Science Foundation (FWF)}}
\maketitle

\begin{abstract}
  In this paper a large class of universal windows for Gabor frames
  (Weyl-Heisenberg frames) is constructed. These windows have the
  fundamental property that every overcritical rectangular lattice
  generates a Gabor frame. Likewise, every undercritical rectangular
  lattice generates a Riesz sequence.
\end{abstract}


%
\IEEEpeerreviewmaketitle

\section{Introduction}
%
%
%
%

\IEEEPARstart{G}{abor} frames and Gabor Riesz sequences arise in many
engineering contexts. The corresponding series expansions can be
interpreted as a sum of local Fourier series, the coefficients carry
simultaneous information about time and frequency. Therefore they are
an appropriate tool whenever simultaneous time-frequency information
is required. Thus Gabor frame expansions are used naturally in speech
processing and in the analysis of music
signals~\cite{dolson86,Picone93signalmodeling}, though often under
names as phase vocoder or lapped Fourier transform. One of the
principal uses of Gabor expansions arises in wireless digital
transmission with OFDM~\cite{MR1955940,HM11,Str01}. A non-exhaustive
list of further applications includes antenna analysis~\cite{LLie02},
the analysis of ultrasound imaging~\cite{MAie02}, of brain stem
responses~\cite{CDSSie09}, and even for the (texture) analysis of
images~\cite{SZie06,zibulski-zeevi98}.

A Gabor family consists of time-frequency shifts of a single window
function $g$ (often called pulse in wireless communications) over a
lattice $\alpha \bZ \times \beta \bZ $. Formally, let $g$ be a
square-integrable function on $\bR $ and $\alpha , \beta >0$ be the
lattice parameters, then the time-frequency shifts are defined as
$M_{\beta n} T_{\alpha m} g(x) = e^{2\pi j \beta n x} g(x-m \alpha )$,
$m,n\in \bZ $. In a concrete problem one is asked to design a pulse
$g$ and choose the time-frequency spacing $\alpha, \beta $ in such a
way that the corresponding set of functions possesses required
properties. Basic requirements on the pulse are often explicit
formulas and good time-frequency concentration, basic requirements on
the set of functions $\cG (g,\alpha ,\beta ) = \{ M_{\beta n
}T_{\alpha m}g : m,n\in \bZ \}$ are its spanning properties. If one
needs a stable expansion of \emph{every} signal with respect to the
time-frequency shifts $M_{\beta n }T_{\alpha m}g$, one has to
construct a Gabor frame (Weyl-Heisenberg frame). In wireless
communications one needs the set $\cG (g,\alpha ,\beta )$ to be
linearly independent and thus has to construct a Gabor Riesz basis for
a proper subspace of $L^2(\bR )$. Currently only few explicit pulse
shapes are used.  The Gaussian function $e^{-a t^2}$ is used in
antenna analysis~\cite{LLie02} or for the compression of EEG
signals~\cite{Avie07}. The rectangular pulse is much in favor in
orthogonal frequency division multiplexing (ODFM) with cyclic
prefix~\cite{MR1955940,HM11}; the raised cosine window is the standard
window in speech processing and the analysis of music signals.  A
representative list of windows for discrete Gabor frames is contained
in Sondergaard's Large Time-Frequency Analysis Toolbox
LTFAT~\cite{ltfat}.  In pulse-shaping
OFDM~\cite{MR1955940,HM11,JWie07,MSGHHie} pulses are constructed to
satisfy certain optimality criteria, but they are usually not given
explicitly.

Once one has chosen a suitable pulse, one needs to determine which
shift parameters generate a Gabor frame or basis. The density theory
of Gabor frames~\cite{daubechies90,heil07} asserts that the \tf\ plane
must be oversampled for $\cG (g,\alpha, \beta )$ to be a frame,
i.e. $\alpha \beta \leq 1$, whereas for a Riesz sequence one needs
undersampling $\alpha \beta \geq 1$. For the case of critical sampling
$\alpha \beta =1$ one can construct orthonormal bases of
time-frequency shifts, but the corresponding pulse always lacks \tf\
concentration~\cite{BHW95}, as can already be seen for the rectangular
pulse.

In many applications it is desirable to operate as close to the
critical density as possible.  For a large  channel capacity a Gabor
Riesz sequence should span a ``large'' subspace of $L^2(\bR )$ and
thus $\alpha \beta = 1+\epsilon $ for some small $\epsilon
>0$. Likewise, to avoid large redundancy of the frame expansion, the
density of a Gabor frame should be $\alpha \beta = 1-\epsilon $ for
small $\epsilon >0$. 

These fundamental requirements pose a difficult and largely unanswered
question.  Once a pulse shape is chosen, it is not at all clear
whether such a choice of lattice parameters close to the critical
density is possible. Until recently the complete set of the lattice
parameters generating a Gabor frame was understood only for a handful
of pulse shapes, namely the Gaussian, the hyperbolic secant, and two
exponential
functions~\cite{janssen96,janssen02c,JS02,lyub92,seip92}. On the other
hand, for the popular rectangular pulse the determination of good
lattice parameters is extremely complicated, as is shown by Janssen's
tie~\cite{janssen02b}.

To this day, it remains a challenging mathematical problem to
determine suitable lattice parameters $\alpha ,\beta $ for which the
corresponding Gabor family $\gab $ is a frame or a Riesz
sequence. Perhaps this lack of theoretical understanding has prevented
the use of other pulses in engineering applications.

In this paper we  propose  totally positive functions as convenient and
universal  windows for Gabor frames. 
The goal  is to offer practitioners of \tfa\ a new, large
class of pulse shapes for which the frame property is easy to
determine for \emph{all} lattice parameters.

Our contribution is based on recent progress in the mathematical
analysis of Gabor frames. In~\cite{grst13}, the following result was
shown. A  function $g$  is said to be totally positive of finite type
$m$ (TPFFT), if its 
Fourier transform $\hat{g}(\xi ) = \int _{-\infty } ^\infty g(x) e^{-2\pi j x\xi }
\, dx$ factors  as follows: let $\delta _i$, $i= 1, \ldots , m$ be a
finite set of nonzero real numbers, then  
$$
\hat{g}(\xi ) = \prod _{i=1}^m (1+2\pi j \delta _i \xi )\inv \, .
$$

\begin{theorem}[~\cite{grst13}]
  \label{thm:grst13-main}
  Let $g $ be a totally positive function of finite
  type $m\ge 2$. Then the Gabor family $\mathcal{G}( g,\alpha, \beta)$
  constitutes a frame if and only if $\alpha \beta < 1$. Moreover the
  Gabor frame possesses a compactly supported, piecewise continuous
  dual window $\gamma$.

By duality, the Gabor family $\gab $  is a Riesz sequence for the
generated subspace, if and only if $\alpha \beta >1$. 
\end{theorem}

This result provides a family of pulse shapes that is parametrized by
a countable number of parameters $\delta _i$. By choosing suitable
$\delta _i$ one can finetune the pulse to one's needs.

We will make this result more useful for applications in signal
processing and for numerical use by deriving similar statements for
Gabor frames for discrete signals, for continuous periodic signals,
and for finite discrete signals. In these cases the Hilbert space is
$\ell ^2(\bZ )$, or $L^2(\bT_K)$, where $K>0$ and \(\bT_K = [0,K]\)
denotes the interval from \(0\) to \(K\) on the real line, or $\bC
^L$, where $L\in\bN$ (instead of $L^2(\bR)$).

The choice of totally positive functions of finite type as a pulse
shape for Gabor frames has several major advantages.
\begin{enumerate}
\item Totally positive functions of finite type (TPFFT) are universal
pulse shapes that generate a Gabor frame whenever the time frequency
plane is oversampled and that generate a Gabor Riesz sequence whenever
the \tf\ plane is undersampled.
\item TPFFTs possess exponential decay. Thus these
pulse shapes can be approximated with high accuracy by a pulse with
compact support. In addition, if the type of $g$ is $m\geq
2$, then $g$ is $m-2$ times continuously differentiable~\cite{MR0047732}.
\item The general theory of Gabor frames guarantees that the Gabor
  frame with a totally positive window function possesses a dual pulse
  with exponential
  decay~\cite{boelcskei-janssen00,delprete,Str01}. The construction
  in~\cite{grst13} yields even a dual pulse with compact support. In
  this paper we refine this construction and present a simple
  algorithm that generates a whole family of dual windows with compact
  support (and increasing smoothness). In contrast to other
  methods~\cite{casazza-christensen00,Str00a,Str01} this algorithm is
  exact and requires only the pseudo-inverse of a finite matrix.
\end{enumerate}

We hope that the new theory of Gabor frames offers a new arsenal of
convenient windows for many applications where the lattice parameters
need to be close to the critical case.

The paper is organized as follows. In Section \ref{sec:not-tools} we
recall the main concepts and necessary results on Gabor frames, we
discuss totally positive functions of finite type and their basic
properties. In Section \ref{sec:main-results} we will state versions
of Theorem \ref{thm:grst13-main} for discrete, periodic and finite
signals, we give an algorithm to compute a dual window \(\gamma\), and
provide explicit formulas for totally positive functions of finite
type and of their Zak transforms. Furthermore we will study the
\emph{critical case} $\alpha \beta =1$.  Section \ref{sec:examples}
contains numerical examples of totally positive functions and a
variety of their dual windows.

\section{Background}
\label{sec:not-tools}
This section is devoted to a brief introduction to Gabor frames and
their discretization. For a thorough introduction see
\cite{MR2428338,MR1843717,MR2744776}.

\subsection{Gabor Frames}
\label{sec:gabor-frames-L2}

Let $g$ be a function in $L^2(\mathbb{R})$. We define the
\emph{translation operator} $T_x$ by $T_xf(t) = f(t - x), t,
x\in\mathbb{R}$, and the \emph{modulation operator} $M_\xi $ by $M_\xi
f(t) = e^{2\pi j \xi t} f(t), \xi\in\mathbb{R}$.  Their composition is
the \emph{time-frequency shift operator}
  \[
  M_\xi T_x f(t) = f(t - x)e^{2 \pi j\xi t}.
  \]
Let $\bT_K=[0,K]$ with $K>0$ and let $L\in\bN$. 
The analogous definitions of the time-frequency shift
operator are given for 
  \begin{itemize}
  \item discrete signals $f\in \ell^2(\bZ)$ by \(M_\xi T_k f(l) = f(l-k)e^{2 \pi j \xi l }\),
     where $k,l\in\bZ$, $\xi \in [0,1)$,
  \item periodic signals $f\in L^2(\bT_K)$ by \(M_{m/K} T_x f(t) :=  
  f((t-x)\, {\rm mod}\, K)e^{2 \pi j m t/K}\),
   where $  t,x\in [0,K]$,  $m\in\bZ$, 
  \item finite discrete signals $f\in\bC^L$ by \(M_{m/L} T_k f(l) :=  
  f((l-k)\, {\rm mod}\, L) e^{2\pi j m l/L}\),  
  where $k,l,m\in \{0,\ldots,L-1\}$.
  \end{itemize}

We first recall the well-known definition of Gabor frames for $\Lt( \bR)$,
see \cite{MR1843717}.

\begin{definition}
  Let $\alpha,\beta \in \mathbb{R^+}$ and $g\in\Lt( \bR)$. 
	The set of time-frequency
  shifts
  \begin{equation}
    \label{eq:gabor-system}
    \mathcal{G}(g,\alpha,\beta) :=\{M_{l\beta} T_{k\alpha}
    g\,|\, k,l \in \bZ\}
  \end{equation}
  is called a \emph{Gabor family}. A Gabor family is a \emph{Gabor
    frame}, or \emph{Weyl-Heisenberg frame}, for $\Lt(\bR)$, if
  there exist constants \(A,B >0\) such that
  \begin{equation}
    \label{eq:frame-condition}
    A\|f\|_2^2 \le \sum_{k,l \in\mathbb{Z}} |\langle f,
    M_{l\beta}T_{k\alpha} g \rangle |^2 \le B\|f\|_2^2\, , \quad \forall f \in
    L^2(\mathbb{R}).
  \end{equation}
  The set of points $\{(k\alpha,l\beta)\in\mathbb{R}^2\,|\, k,l \in
  \mathbb{Z}\}$ is  a lattice with \emph{density} $(\alpha\beta )\inv $. The
  constants  \(A,B\) in~\eqref{eq:frame-condition}  are called the
  \emph{frame bounds}.

The Gabor family $\gab $ is a (Gabor) \emph{Riesz sequence}, if there
exist constants $A',B'>0$, such that 
\begin{equation}
  \label{eq:c2}
  A' \|c\|_2^2 \leq \left\|\sum _{k,l\in \mathbb{Z}}  c_{kl}
  M_{l\beta}T_{k\alpha} g  \right\|_2^2 \leq B' \|c\|_2^2 \, , \quad \forall c\in
  \ell ^2(\mathbb{Z}^2) \, .
\end{equation}
\end{definition}

The usefulness of Gabor frames stems from the
  fact that they allow for a basis like expansion of functions in
  \(\Lt(\bR)\).

  \begin{proposition}
    \label{prop:gabor-expansion}
  (~\cite{MR1843717})
  Let \(\gab\) be a Gabor frame for \(\Lt(\bR)\), then there exists a
  \emph{dual window} \(\gamma \in \Lt(\bR)\), such that every \(f\in
  \Lt(\bR)\) possesses the expansions
  \begin{align*}
    f &= \sum_{k,l\in\bZ} \langle f, M_{l\beta }T_{k\alpha }g \rangle
    M_{l\beta } T_{k\alpha }\gamma \\
    &= \sum_{k,n\in\bZ} \langle f, M_{l\beta }T_{k\alpha }\gamma \rangle
    M_{l\beta } T_{k\alpha }g,
  \end{align*}
 and the set \(\cG(\gamma, \alpha, \beta)\) is also a frame for $L^2(\bR)$. 
 This frame is called a \emph{dual
    frame} of \(\gab\).
  \end{proposition}

We note that the dual window is not unique, a  formula for all
possible dual windows can be found in~\cite{MR2428338,MR1843717}. The
non-uniqueness will offer us some freedom to design a class of windows
with compact support in Section~\ref{sec:algo-gamma}.

Gabor frames and Gabor Riesz sequences are dual to each other~\cite{janssen95,ron-shen97}. In
fact, the Gabor family $\cG (g,\alpha , \beta )$ is a frame for
$L^2(\bR)$, if and only if $\cG (g, \tfrac{1}{\beta},
\tfrac{1}{\alpha})$ is a Riesz sequence in $L^2(\bR )$.  This
connection has been widely used in wireless communications~\cite{HM11}.

The definition of a Gabor family easily carries over to the setting of
$\ltZ$, $\Lt(\bT_K)$ and $\bC^L$.
\begin{itemize}
\item For $g\in \ell^2(\bZ)$ we fix $\alpha\in\bN$,
  $\beta=\frac{1}{M}$ with $M\in \bN$ and let
  \begin{equation}
      \label{eq:gab-sys-l2}
      \cG\left(g,\alpha,\beta \right)= \{M_{l\beta}T_{k\alpha}g\,|\,
      k\in\bZ,l=0, \ldots,M-1\}.
    \end{equation}
  \item For $g\in \Lt(\bT_K)$ we fix $\alpha=\frac{K}{N}$ with $N\in
    \bN$, $\beta=\frac{p}{K}$ with $p\in \bN$ and let
    \begin{equation}
      \label{eq:gab-sys-TK}
      \cG\left(g,\alpha,\beta \right)= \{M_{l\beta}T_{k\alpha}g\,|\,
      k=0, \ldots,N-1,~l\in\bZ\}.
    \end{equation}
  \item For $g\in \bC^L$ we fix $\alpha=\frac{L}{N}$ with $N\in \bN$,
    $\beta=\frac{1}{M}$ with $M\in \bN$, where we suppose that
    $\frac{L}{N},\frac{L}{M}\in \bN$ and let
    \begin{equation}
      \label{eq:gab-sys-CK}
      \cG\left(g,\alpha,\beta \right)= \{M_{l\beta}T_{k\alpha}g\,|\,
      k=0, \ldots,N-1,~l=0,\ldots,M-1\}.
    \end{equation}
  \end{itemize}
  The analogous definition of Gabor frames for \(\lt(\bZ),
  \Lt(\bT_K)\) and \(\bC^L\) requires that the inequalities in
  \eqref{eq:frame-condition} hold for all $f$ in the respective
  Hilbert space, and the summation extends over all index pairs
  $(k,l)$ that are relevant for the Gabor family instead of
  $\bZ\times\bZ$.  Analogous results to Proposition
  \ref{prop:gabor-expansion} hold for Gabor frames for \(\lt(\bZ),
  \Lt(\bT_K)\) and \(\bC^L\).

\subsection{Sampling and Periodization of Gabor Frames}
\label{sec:samp-per-gf}

First, we introduce the sampling and periodization operators.  Let \(g
\in  \Lt(\bR)\) and let \(h\in\bR\) with \(h>0\).
We assume that every point $t=hk$ with $k\in\bZ$ is a 
Lebesgue point of \(g\), i.e., evaluation of $g$ at $t$ is defined because
\[
   \lim_{\epsilon\to 0} \frac{1}{2\epsilon}
	\int_{-\epsilon}^\epsilon |g(x+u)-g(x)|\,du=0.
\]
 Then
\begin{equation}
  \label{eq:sampling-op}
  S_h g := \left(g(hk): k\in\bZ\right)
\end{equation}
defines the \emph{sampling operator} with step-size \(h\). 
It was shown in~\cite{MR1491936} and~\cite{MR2349904} that $g\in
L^2(\bR )$ and 
$S_h g\in \lt(\bZ)$ holds
under the slightly stronger condition
  \begin{equation}
    \label{eq:condition-R}
    \lim_{\varepsilon \rightarrow 0} \sum_{k=-\infty}^\infty
    \frac{1}{2\varepsilon} \int_{-\varepsilon}^{ 
      \varepsilon} |g(kh+u)-g(kh)|^2 du = 0.
  \end{equation}
This condition is clearly satisfied, if $g$ is 
piecewise differentiable and its derivative decays 
exponentially, as will be the case for all $g$ considered
in this article.

For periodization with period
\(K\in\bR\), $K>0$, we assume $g\in \Lo(\bR)$ and let
\begin{equation}
  \label{eq:periodization-op}
  \mathcal{P}_Kg(x):= \sum_{k\in\mathbb{Z}} g(x- kK), \quad x\in\bT_K.
\end{equation}
Then  $P_Kg\in \Lo(\bT_K)$, and if $g$ is 
piecewise continuous and  decays 
exponentially, then $P_Kg\in \Lt(\bT _K)$. 
We call $P_K$ the \emph{periodization operator} with period $K$. 
With $K\in\bN$, the 
periodization operator $P_K$ is defined analogously on $\ell^1(\bZ)$;
its image space is isomorphic to $\bC^K$.

The family of window functions $g$ considered in this article
is 
invariant under dilation. Therefore, it is no restriction for us 
to let \(h=1\) in the sequel
and write \(Sg\) instead of \(S_1g\).

It was shown in~\cite{MR1491936} and~\cite{MR2349904} that Gabor frames for
\(\lt(\bZ)\), \(\Lt(\bT_K)\) and \(\bC^K\) can be derived from Gabor frames for
\(\Lt(\bR)\) by sampling and periodization. The precise statements are
as follows.

\begin{proposition}
  \label{prop:janssen-samp}
  Let $\alpha\in\mathbb{N}$, $\beta=1/M$ with $M\in\bN$, and $g\in L^2(\mathbb{R})$ such that
   $\mathcal{G}(g,\alpha,\beta)$ is a Gabor frame
  for \(\Lt(\bR)\) with frame bounds $A,B>0$.  If $g$ satisfies the condition 
	\eqref{eq:condition-R},
  then
  \begin{itemize}
  \item $\mathcal{G}(Sg,\alpha,\beta)$ is a Gabor frame for \(\lt(\bZ)\) with
    the same frame bounds,
  \item \(\cG (P_K g, \alpha,\beta)\) is a Gabor frame for \(\Lt(\bT_K)\)
    with the same frame bounds, for any period $K\in\bN$ with
		$\frac{K}{\alpha}\in\bN$ and $\frac{K}{M}\in\bN$,
  \item $\mathcal{G}(\mathcal{P}_K S g, \alpha,\beta)$ is a Gabor frame for
    $\mathbb{C}^K$ with the same frame bounds, for any period $K\in\bN$ as above.
  \end{itemize}
Moreover, if $\mathcal{G}(\gamma,\alpha,\beta)$ is a dual Gabor frame of 
$\mathcal{G}(g,\alpha,\beta)$ in $\Lt(\bR)$ 
and $\gamma$ satisfies \eqref{eq:condition-R},
then $S\gamma$, $P_K\gamma$ and $P_KS\gamma$ are window functions of
a dual Gabor frame after sampling and/or periodization.
\end{proposition}

In this proposition, the statement about frame bounds does not mean that
the optimal frame bounds remain the same. In many cases, tighter bounds
than $A,B$ exist after periodization or sampling. 

\subsection{Density and Zak transform}
The  lattice $\alpha\bZ \times \beta\bZ$ of time-frequency shifts 
of $g\in\Lt(\bR)$
must satisfy a density criterion, if $\cG(g,\alpha,\beta)$ defines a Gabor frame
for $\Lt(\bR)$, see \cite{heil07}.
 Analogous results 
for Gabor frames for \(\lt(\bZ), \Lt(\bT_K)\) and \(\bC^L\) are
easier to obtain, see, e.g.,  \cite{fekolu09}.

\begin{theorem} Let $\cH$ be one of the spaces \(\Lt(\bR),
\lt(\bZ), \Lt(\bT_K)\) or \(\bC^L\), and $g\in\cH$.
If $\cG(g,\alpha,\beta)$ is a Gabor frame for $\cH$, then
$\alpha\beta\le 1$. 
\end{theorem}

For the case $\alpha\beta=1$, which is called the {\em critical
density}, the frame condition \eqref{eq:frame-condition} can be checked by
inspecting the  \emph{Zak transform} (cf.~\cite{564174,MR1843717,he89,janssen-zak88}) of $g$.
\begin{definition}
  Let \(f\in\Lt(\bR)\), \(\alpha > 0\). The \emph{Zak transform}
  \(Z_\alpha f\) of \(f\) is defined as
  \begin{equation}
    \label{eq:zak-transform-continuous}
    Z_\alpha f(x,\xi) = \sum_{k\in \bZ} f(x - \alpha k) e^{2 \pi j
      \alpha k \xi},\qquad x,\xi\in\bR.
  \end{equation}
\end{definition}
The Zak transform is \emph{quasiperiodic}, i.e.,
\begin{align}
  \label{eq:zak-cont-per-rel-1}
  Z_\alpha f(x,\xi+n/\alpha) &= Z_\alpha f(x,\xi),\\
  \label{eq:zak-cont-per-rel-2}
  Z_\alpha f(x+\alpha n,\xi) &= e^{2 \pi j \alpha n \xi} Z_\alpha
  f(x,\xi), \quad n \in \bZ.
\end{align}
The quasiperiodicity implies that the Zak transform on \(\bR^2\) is
determined by its values on the rectangle \([0,\alpha) \times
[0,1/\alpha)\).

The subsequent theorem gives a criterion to check whether the frame condition 
\eqref{eq:frame-condition} is satisfied at the critical density $\alpha\beta=1$.
In particular, we will see that the frame property of a
Gabor set is completely determined by the behaviour of the Zak
transform of the window \(g\).
\begin{theorem}{(cf.~\cite{564174,MR2428338,MR1843717,he89})}
  \label{lem:zak-ineq}
  Let \(g\in\Lt(\bR)\) and \(\alpha \beta = 1\).
  \begin{itemize}
  \item[(a)] \(\cG(g,\alpha,\beta)\) is a
    frame for \(\Lt(\bR)\) with frame bounds \(A,B\) if and only if
    \begin{equation}
      \label{eq:zak-ineq-cont}
      0<A\le \absl Z_\alpha g(x,\xi)\absr^2 \le B <\infty
    \end{equation}
		for almost all $(x,\xi)\in [0,\alpha)\times[0,1/\alpha)$.
  \item[(b)] Assume that \(g\) satisfies (\ref{eq:condition-R}) and 
    \(\alpha=M \in\bN\), $\beta=1/M$. Then \(\cG(Sg,\alpha,\beta)\) is a frame for \(\lt(\bZ)\) with
    frame bounds \(A,B\), if and only if
    \begin{equation}
      \label{eq:zak-ineq-lt}
      0<A\le |Z_\alpha g(k,\xi)|^2\le B < \infty
    \end{equation}
		for $k\in \{0,\ldots,
      M-1\}$ and almost all $\xi \in [0,1/M)$.
  \item[(c)] Assume that \(g\) satisfies (\ref{eq:condition-R}) and 
    \(\alpha=M\in\bN\), $\beta=1/M$. Let $K\in\bN$ such that $K/M\in\bN$.
		 Then
    \(\cG(P_Kg,\alpha,\beta)\) is a frame for \(\Lt(\bT_K)\) with frame bounds
    \(A,B\), if and only if
    \begin{equation}
      \label{eq:zak-ineq-TK}
      0<A\le \left|Z_\alpha g\left(x,\frac{l}{K}\right)\right|^2\le B < \infty
    \end{equation}
		for almost all $x\in [0,M)$  and all $l \in \{0,1,\ldots,\frac{K}{M}-1\}$.
  \item[(d)] Assume that \(g\) satisfies (\ref{eq:condition-R}) and
      \(\alpha=M\in\bN\), $\beta=1/M$. Let $K\in\bN$ such that $K/M\in\bN$. 
			Then \(\cG(P_K Sg,\alpha,\beta)\) is a frame for \(\bC^K\) with frame bounds \(A,B\), if
    and only if
    \begin{equation}
      \label{eq:zak-ineq-CK}
      0<A\le \left|Z_\alpha g\left(k,\frac{l}{K}\right)\right|^2\le B < \infty
    \end{equation}
		for all $k\in \{0,\ldots,
      M-1\}$ and $l \in \{0,1,\ldots, \frac{K}{M}-1\}$.
  \end{itemize}
\end{theorem}

The Balian-Low theorem \cite{BHW95} states that
windows $g\in\Lt(\bR)$ with the property $g,\hg\in\Lo(\bR)$ 
cannot define a Gabor frame for $\Lt(\bR)$ at the critical density $\beta=1/\alpha$.
This was proved in \cite{daubechies90} by applying the following result in connection with
Theorem~\ref{lem:zak-ineq}(a).

\begin{lemma}
  \label{lem:zak-zeros-cont}
  Let $\alpha>0$, \(g \in \Lt(\bR)\). If \(Z_\alpha g\) is continuous,
  then it has a zero in its domain \([0,\alpha) \times [0,1/\alpha)\).
\end{lemma}

A more precise statement about the location of some zero of $Z_\alpha g$
was given in \cite{janssen02b} for the case of 
an even function $g$.

\begin{lemma}
  \label{lem:zeros-of-zak}
  Let $\alpha>0$, \(g\in\Lt(\bR)\) be even and $Z_\alpha g$ be continuous. Then $ Z_\alpha
  g(\frac{\alpha}{2},\frac{1}{2\alpha})=0$. 
	
	Moreover, assume $\alpha=M\in\bN$ and let $\beta=1/M$ and $K\in\bN$ is such that $K/M\in\bN$. 
  \begin{itemize}
  \item \(\cG( Sg,\alpha,\beta)\) is not a Gabor frame for $\lt(\bZ)$ if $M$ is even.
  \item \(\cG( P_Kg,\alpha,\beta)\) is not a Gabor frame for $\Lt(\cT_K)$ if $K/M$ is even.
  \item \(\cG( P_KSg,\alpha,\beta)\) is not a Gabor frame for $\bC^K$ if $M$ and $K/M$ are even.
  \end{itemize}
\end{lemma}

\begin{IEEEproof}
The zero at $(\frac{\alpha}{2},\frac{1}{2\alpha})$ 
of the Zak transform $Z_\alpha g$ of an  even continuous 
 function $g$
was already described in \cite{janssen02b}. 
If $M$ is even, the point $(k,\xi)=(\frac{M}{2},\frac{1}{2M})$
is in the domain of $Z_\alpha g$ in part (b) of Theorem
\ref{lem:zak-ineq}. Therefore, no positive lower frame bound 
exists for the Gabor family $\cG(Sg,\alpha,\beta)$. An analogous
argument is used to show that \(\cG( P_Kg,\alpha,\beta)\)
and \(\cG( P_KSg,\alpha,\beta)\) have no positive lower frame bound,
if the conditions on $M$ and $K$ are satisfied.
\end{IEEEproof}

\subsection{Totally Positive Functions}
\label{sec:tot-pos-fun}

\begin{definition}
  A non-constant function $g\in L^1(\mathbb{R})$ is said to be
  \emph{totally positive} if for every two sequences
  \begin{align}
    x_1 < x_2 &< \cdots < x_N, \notag\\
    y_1 < y_2 &< \cdots < y_N
  \end{align}
  of real numbers the inequality
  \begin{equation*}
    \det (g(x_j - y_k))_{1\le j,k \le N} \ge 0
  \end{equation*}
  holds.
\end{definition}

In~\cite{MR0047732} Schoenberg showed that the total positivity of a
function $g\in L^1(\mathbb{R})$ is equivalent to a simple
factorization of its Laplace transform. In this paper we will study
the special case of \emph{totally positive functions of finite
  type}. Instead of the Laplace transform we use the Fourier transform
in the following definition.
\begin{definition}
  A function $g\in L^1(\mathbb{R})$ is \emph{totally positive of
    finite type $m \in \bN$}, if its Fourier transform is given by
  \begin{equation}
  \label{eq:finite-factorization}
    \hat{g}(\xi) = C\prod_{k = 1}^m (1 + 2 \pi j\delta_k \xi)^{-1}
  \end{equation}
where $C>0$ and $\delta_k\ne 0$ are real numbers.
\end{definition}

\begin{example} In the following examples we let $C=1$ in
  \eqref{eq:finite-factorization} and we make use of the Heaviside
  function
  \[
  h(x)=\begin{cases} 1,&x>0,\\
    \frac{1}{2},&x=0,\\  0,&x<0.\end{cases}
  \]
  The two sided exponential function $g(x) = \frac{1}{2}e^{-|x|}$ is
  totally positive of finite type \(m=2\), with parameters $\delta_{1}
  =1, \delta _2=- 1$.  Also of type $m=2$ are the functions $g(x)
  =\frac{ab}{a+b}( e^{ax} h(-x) + e^{-bx} h(x))$ for $a,b>0$, with
  parameters $\delta_1=-1/a$ and $\delta_2=1/b$, and $g(x)
  =\frac{ab}{b-a}((e^{-ax}-e^{-bx})h(x)$ for $b>a>0$, with
  $\delta_1=1/a$ and $\delta_2=1/b$. An example of a totally postive
  function of type \(m=r\in\bN\) is $g(x) = e^{-x}
  \frac{x^{r-1}}{(r-1)!} h(x)$, with parameters $\delta_1=\cdots
  =\delta_r=1$. In Theorem \ref{thm:explicit-formula-continuous} we
  will give a direct formula for totally positive functions of finite
  type, without reference to their Fourier transform. This makes their
  evaluation very simple. By their definition, the dilation and the
  convolution of totally positive functions of finite type is again
  totally positive of finite type. Note that the Gaussian $g(x) =
  e^{-\pi x^2}$ is totally positive, but since its Fourier transform
  is again a Gaussian it is \emph{not} of finite type.
\end{example}

We state two basic properties of totally positive functions which will be
needed later on. For a proof see~\cite{grst13} and~\cite[p.339]{MR0047732}.

\begin{proposition}(Properties of totally positive functions)
  \label{prop:tpfft-properties}
  \begin{enumerate}
  \item Every totally positive function $g\in \Lo(\bR)$ has 
	exponential decay.
  \item Every totally positive function $g$ of finite type $m\ge 2$
	is continuous.
  \end{enumerate}
	Consequently, every totally positive function $g$ of finite type 
	satisfies condition \eqref{eq:condition-R}, and the Zak transform of 
	every totally positive function of finite type $m\ge 2$ is continuous.
\end{proposition}


\section{Main Results}
\label{sec:main-results}

In this section we will state a version of Theorem
\ref{thm:grst13-main} for the case of discrete signals $f \in
\ell^2(\mathbb{Z})$, periodic signals \(f\in \Lt(\bT_K)\), and the
case of finite signals $f\in\mathbb{C}^K$.  We will give explicit
formulas for the respective settings and treat the case of the {\em
  critical density} where \((\alpha,\beta)=(M,1/M)\) with $M\in \bN$
in detail. This case differs significantly from the setting of Gabor
frames of $\Lt(\bR)$, as there exist Gabor frames for $\ell ^2(\bZ )$
and $L^2(\bT )$ at the critical density which are well localized in
time and frequency.  We will also describe an efficient algorithm for
the computation of a dual window $\gamma$ in all three cases.

\subsection{TPFFTs and their Zak transform}
\label{sec:exp-forms}

First we will present explicit formulas for the computation of
totally positive functions of finite type and their Zak transforms in
the continuous setting. 

\begin{theorem}
\label{thm:explicit-formula-continuous}
Let $g$ be a totally positive function of finite type $m\ge 2$ with
Fourier transform
  \begin{equation}
    \label{eq:tpfft-fourier-transform}
    \hg(\xi) = \prod_{k = 1}^m(1+2 \pi j\delta_k\xi)\inv
  \end{equation}
  and $\delta_k \in \mathbb{R} \setminus \{0\}$, and suppose
	$\delta_i
  \neq \delta_k$ for $i \neq k$. Then $g$ is given by
\begin{equation}
  \label{eq:tpfft-explicit-formula}
  g(x) = \sum_{i=1}^m \left(
    \frac{1}{|\delta_i|} e^{-\frac{x}{\delta_i}} 
		h(x \delta_i) \prod_{k = 1,\, k\neq i}^m
    \left(1-\frac{\delta_k}{\delta_i}\right)^{-1}\right).
\end{equation}
\end{theorem}
\begin{IEEEproof}
  We use the partial fraction decomposition of $\hg$,
  \begin{equation}
    \label{eq:par-frac-decomp}
  \hg(\xi) = \prod_{k=1}^m (1+2 \pi j\delta_k\xi)\inv  =
  \sum_{k=1}^m \frac{C_k}{1+2 \pi j \delta_k\xi}.
  \end{equation}
  To find $C_i$ for $i\in\{1,\ldots,m\}$, we multiply
  (\ref{eq:par-frac-decomp}) by $(1+2 \pi j \delta_{i} \xi)$ and
  substitute $\xi = -(2 \pi j\delta_{i})\inv$, then 
  \begin{equation}
    \label{eq:C_j}
    C_{i} = \prod_{k=1,k\neq i}^m
    \left(1-\frac{\delta_k}{\delta_{i}}\right)\inv.
  \end{equation}
  The $i$-th summand $\hat s_i(\xi)$ 
	in \eqref{eq:par-frac-decomp} is in $\Lt(\bR)$,
	and its inverse Fourier transform is 
	\[
	  s_i(x)= 
    \frac{C_i}{|\delta_i|}\, e^{-\frac{x}{\delta_i}} \,
		h(x \delta_i).
	\]
	This shows that $g(x)$ has the form \eqref{eq:tpfft-explicit-formula}
	for all $x\ne 0$, since the summands $s_i$ are continuous in $\bR\setminus\{0\}$.
	Continuity of $g$ in $x=0$, as stated in Proposition \ref{prop:tpfft-properties}, 
	implies 
\[	 
    g(0)=\lim_{x\searrow 0}g(x)=\sum_{i=1,\delta_i>0}^m 
    \frac{C_i}{\delta_i},\qquad g(0) = \lim_{x\nearrow 0}g(x) 
		= \sum_{i=1,\delta_i<0}^m 
    \frac{C_i}{|\delta_i|} .
\]
Therefore, we conclude that 	
\[
   g(0)=\frac{1}{2}\left(\sum_{i=1,\delta_i>0}^m 
    \frac{C_i}{\delta_i} + \sum_{i=1,\delta_i<0}^m 
    \frac{C_i}{|\delta_i|}\right) =h(0) \sum_{i=1}^m 
    \frac{C_i}{|\delta_i|},
\]
which agrees with the explicit form in \eqref{eq:tpfft-explicit-formula}.
\end{IEEEproof}

\begin{remark}
  Similar expressions for TPFFTs can be derived, if $\hg$ has poles of
  higher multiplicity, but the formulas quickly become more
  involved. An extension of the result in Theorem
  \ref{thm:explicit-formula-continuous} to totally positive functions
  \(g\), where \(\hat{g}\) has infinitely many poles, is given in
  \cite[Theorem 4]{MR0047732}.
\end{remark}
Formula (\ref{eq:tpfft-explicit-formula}) can be used to derive an
explicit expression for the Zak transform of totally positive
functions of finite type.
\begin{corollary}\label{cor:tpfft-zak}
  Let \(g\) be a totally positive function of finite type \(m\ge 2\) with 
	Fourier transform \eqref{eq:tpfft-fourier-transform} and
  \(\delta_k\) as in Theorem \ref{thm:explicit-formula-continuous}. For $\alpha>0$ 
	and
  \(x \in [0,\alpha)$, $\xi \in [0,1/\alpha)\), we have
  \begin{equation}
    \label{eq:zak-explicit-formula}
    Z_\alpha g(x,\xi) = \sum_{i=1}^m \frac{1}{\delta_i}\,
    \frac{e^{-x /\delta_i}}{1- e^{-\alpha (1/\delta_i + 2 \pi j
        \xi)}}
    \cdot \prod_{k = 1, k\neq i}^m \left(1- \frac{\delta_k}{\delta_i}
    \right)^{-1}.
  \end{equation}
\end{corollary}
\begin{IEEEproof}
  The continuity of $Z_\alpha g$ in Proposition \ref{prop:tpfft-properties}
	allows us to ignore $x=0$
	and  assume $x\in(0,\alpha)$ and $\xi\in\bR$.
	We consider each summand 
	\[
	  s_i(x)= 
    \frac{C_i }{|\delta_i|} e^{-\frac{x}{\delta_i}} 
		h(x \delta_i)\]
of $g$ in \eqref{eq:tpfft-explicit-formula} with $C_i$ in \eqref{eq:C_j} separately.
For $\delta_i>0$ we obtain, by the formula for the geometric series,
  \begin{align}
    \label{eq:zak-explicit-formula-posdj}
    Z_\alpha s_i(x,\xi) &= \frac{C_i}{\delta_i} \sum_{l=-\infty}^\infty 
   e^{-(x-\alpha l )/\delta_i}
    h((x-\alpha l)\delta_i)
    e^{2\pi j \alpha l \xi} \notag \\
    &= \frac{C_i}{\delta_i} \,e^{-x/\delta_i}
    \sum_{l=0}^\infty e^{-\alpha l (1/\delta_i + 2\pi j \xi)}
     \notag\\
    &= \frac{C_i}{\delta_i} \,
    \frac{e^{-x/\delta_i}}{1-e^{-\alpha (1/\delta_i + 2 \pi j
        \xi)}}.
  \end{align}
  A similar computation for \(\delta_i<0\) yields
  \begin{align}
    \label{eq:zak-explicit-formula-negdj}
    Z_\alpha s_i(x,\xi) &= -\frac{C_i}{\delta_i} \sum_{l=-\infty}^\infty 
   e^{-(x-\alpha l )/\delta_i}
    h((x-\alpha l)\delta_i)
    e^{2\pi j \alpha l \xi} \notag \\
    &= -\frac{C_i}{\delta_i} \,e^{-x/\delta_i}
    \sum_{l=1}^\infty e^{\alpha l (1/\delta_i + 2\pi j \xi)}
     \notag\\
    &= -\frac{C_i}{\delta_i} \,
    \frac{e^{-x/\delta_i}e^{\alpha  (1/\delta_i + 2\pi j \xi)}}
		{1-e^{\alpha (1/\delta_i + 2 \pi j\xi)}}\notag\\
		&=\frac{C_i}{\delta_i} \,
    \frac{e^{-x/\delta_i}}
		{1-e^{-\alpha (1/\delta_i + 2 \pi j\xi)}}.
  \end{align}
This completes the proof of \eqref{eq:zak-explicit-formula}.
\end{IEEEproof}

\begin{remark}\label{rem:divdiff}
The right-hand side in \eqref{eq:zak-explicit-formula} can 
be interpreted as the expanded form of a divided 
difference of order $m-1$ with knots $a_i:=1/\delta_i$, $1\le i\le m$,
of the function 
\begin{equation}\label{eq:rx}
   r_{x,\xi}(y)=(-1)^{m-1}\left(\prod_{i=1}^m a_i\right) \frac{e^{-xy}}{1-e^{-\alpha(y+2\pi j\xi)}}. 
\end{equation}
Indeed,
the divided difference with knots $a_i$ has the expanded form
\[
   [a_1,\ldots,a_m\mid r_{x,\xi}] = \sum_{i=1}^m r_{x,\xi}(a_i) 
	\prod_{k = 1, k\neq i}^m \left(a_i- a_k \right)^{-1},
\]
which coincides with the right-hand side in 
\eqref{eq:zak-explicit-formula} by straightforward
computation. This allows  us to extend the result of 
Corollary \ref{cor:tpfft-zak} to totally positive functions of 
finite type with parameters $\delta_i$ among which several may coincide,
by taking the divided difference of $r_{x,\xi}$ with corresponding multiple knots
$a_i=1/\delta_i$, $1\le i\le m$. By this, we obtain the representation of $Z_\alpha g$
for every TPFFT
\begin{equation}\label{eq:tpfft-zak-general}
 Z_\alpha g(x,\xi) =[a_1,\ldots,a_m\mid r_{x,\xi}],
\end{equation}                                                  
where \(x \in [0,\alpha)$, $\xi \in [0,1/\alpha)\) and $r_{x,\xi}$ is given in 
\eqref{eq:rx}. 
\end{remark}

\subsection{Gabor Frames of TPFFTs on \(\lt(\bZ)\)}
\label{sec:gf-tpfft-ltz}

In this section we will state the analog of Theorem
\ref{thm:grst13-main} for \(\lt(\bZ)\).
\begin{theorem}
  \label{thm:grst13-main-lp}
  Let $g$ be a totally positive function of finite type \(m\ge 2\),
  and let $a,M \in \mathbb{N}$. If $\frac{a}{M} < 1$, then $
  \mathcal{G}(Sg,a,\frac{1}{M})$ is a frame for $\ell^2 (\mathbb{Z})$.
  Furthermore there exists a finitely supported dual window $\gamma\in\lt(\bZ)$
  that can be calculated by Algorithm \ref{algo:gamma}.

By duality, if $\frac{a}{M}>1$, then  $\cG (Sg, a, \frac{1}{M})$ is a Riesz sequence in $\ell
^2(\bZ )$. 
\end{theorem}
\begin{IEEEproof}
  Since every totally positive function of finite type has exponential decay and a
  totally positive function of finite type \(m\ge 2\) is continuous by
  Proposition \ref{prop:tpfft-properties}, \(g\) satisfies
  (\ref{eq:condition-R}). Since \( \mathcal{G}(g,a,1/M)\) is a Gabor
  frame for \(\Lt(\bR)\) by Theorem \ref{thm:grst13-main}, the first
  statement of Proposition \ref{prop:janssen-samp} implies that
  \(\mathcal{G}(Sg,a,1/M)\) is a frame for \(\lt(\bZ)\). The existence
  of a dual window \(\gamma\) with finite support is proved in the
  appendix.
\end{IEEEproof}

\subsection{Gabor Frames of TPFFTs on \(\Lt (\bT_K)\)}
\label{sec:gf-tpfft-LTT}

We next formulate the main result for continuous periodic signals in
\(\Lt(\bT_K)\) where $K>0$. Note that periodization of the TPFFT  $g$ 
is defined by
\[
    (\mathcal{P}_K g)(x) = \sum_{k\in\bZ}g(x-kK)=Z_Kg (x,0)\, .
\]
If all parameters $\delta_i\in \bR\setminus\{0\}$ are distinct, then,
by Corollary \ref{cor:tpfft-zak}, 
\begin{equation}
    \label{eq:div-diff-samp-per-I}
     (\mathcal{P}_K g)(x) = 
		\sum_{i=1}^m \frac{1}{\delta_i}\,
    \frac{e^{-x /\delta_i}}{1- e^{- K/\delta_i }}
    \cdot \prod_{k = 1, k\neq i}^m \left(1- \frac{\delta_k}{\delta_i}
    \right)^{-1}
  \end{equation}
for all $x\in\bT_K$.
If multiple entries $\delta_i$ occur, we can switch to the representation 
of $Z_Kg(x,0)$ by a
divided difference of order $m-1$ with knots $a_i=1/\delta_i$, as in Remark
\ref{rem:divdiff}. This gives
\begin{equation}
    \label{eq:div-diff-periodic}
     (\mathcal{P}_K g)(x) = [a_1,\ldots,a_m\mid r_{x,0}]
\end{equation}
where 
\[  r_{x,0}(y)=	(-1)^{m-1}\left(\prod_{i=1}^m a_i\right) \frac{e^{-xy}}{1-e^{-Ky}}. 
\]
	
\begin{theorem}
  Let $g$ be a totally positive function of finite type \(m\ge 2\) and
  let $a,K,M \in \mathbb{N}$, such that $K/a\in\bN$ and $K/M\in\bN$. If
  $\frac{a}{M} < 1$,  then $ \mathcal{G}(P_K g,a,1/M)$ is a
  frame for $\Lt(\bT_K)$.  Furthermore there exists a dual window $P_K
  \gamma$, which is obtained by periodizing the dual window computed
  by Algorithm \ref{algo:gamma}.

By duality, if $\frac{a}{M}>1$, then $ \mathcal{G}(P_K g,a,1/M)$ is a
Riesz sequence for $L^2(\bT_K)$. 
\end{theorem}
\begin{IEEEproof}
  The proof is the same as the proof of Theorem
  \ref{thm:grst13-main-lp}, except that we now apply the second
  statement of Proposition \ref{prop:janssen-samp}.
\end{IEEEproof}

\subsection{Gabor Frames of TPFFTs on \(\bC^K\)}
\label{sec:gf-tppft-CK}

The analog of Theorem \ref{thm:grst13-main} for finite, discrete
signals in \(\bC^K\) is as follows.
\begin{theorem}
  Let $g$ be a totally positive function of finite type \(m\ge 2\) and
  $a,K,M \in \mathbb{N}$ such that \(K/a\in\bN\) and \(K/M \in \bN\).
  If $\frac{a}{M} < 1$, then $ \mathcal{G}(P_K Sg,a,\frac{1}{M})$ is a
  frame for $\mathbb{C}^K$.  Furthermore a dual window $P_K S \gamma$
  can be obtained by periodization and sampling the (continuous) dual
  window computed by Algorithm \ref{algo:gamma}.
\end{theorem}
\begin{IEEEproof}
  Again the proof is the same as the proof of Theorem
  \ref{thm:grst13-main-lp}. This time we use the third statement of
  Proposition \ref{prop:janssen-samp}.
\end{IEEEproof}

\begin{remark}
  In applications of digital signal processing, the sampling step-size
  $h=1/N$ (in seconds), where $N\in\bN$, is often fixed by physical
  measurements or devices for transmission.  Likewise, the width of
  the window $g$ (such as the halfband-width of the Gaussian window)
  is also fixed beforehand. In order to avoid unnecessary scaling of
  $g$, one uses the periodization of the sampled window at the proper
  step-size, that is
\[
    (\mathcal{P}_K S_{1/N}g)\left(\frac{l}{N}\right) = \sum_{k\in\bZ}g\left(\frac{l}{N}-kK\right)=
		Z_Kg \left(\frac{l}{N},0\right)
\]
with $l\in\{0,\ldots,NK-1\}$ and period of length $K\in\bN$. By additional 
normalization in $\bC^{NK}$, and for a totally positive function
$g$ of finite type $m\ge 2$ with distinct nonzero parameters $\delta_k$, we define 
\[
    (\mathcal{Q}_{K,N} g)(l) = N^{-1/2}
		Z_Kg \left(\frac{l}{N},0\right)=
			N^{-1/2}	\sum_{i=1}^m \frac{1}{\delta_i}\,
    \frac{e^{-l /(N\delta_i)}}{1- e^{- K/\delta_i }}
    \cdot \prod_{k = 1, k\neq i}^m \left(1- \frac{\delta_k}{\delta_i}
    \right)^{-1}
\]
for all $l\in \{0,\ldots,NK-1\}$. This gives a practical way to
quickly generate sampled and periodized windows for Gabor frames. One
just needs to choose pairwise distinct values $\delta_i$ to compute a
vector $g$ of length $KN$ according to sampling step-size $h=1/N$ or
$h=1$ and period of length $K$.  Then $\mathcal{G}(g,a,1/M)$ is a
Gabor frame for $\mathbb(\bC^{KN})$ for arbitrary $a,M \in \mathbb{N}$
with $a/M<1$, provided that $K/a$ and $K/M$ are integers.
\end{remark}

\subsection{The critical density}
\label{sec:crit-cas-aM1}

In this section we will study Gabor families $\cG(Sg,\alpha,\beta)$,
$\cG(P_Kg,\alpha,\beta)$, and $\cG(P_K Sg,\alpha,\beta)$ at the
critical density \(\alpha=M\in\bN\), $\beta=1/M$. Note that for a
totally positive function $g$ of finite type $m\ge 2$, Theorem
\ref{thm:grst13-main} says that \(\alpha\beta<1\) is equivalent to
\(\cG(g,\alpha,\beta)\) being a frame for \(\Lt(\bR)\). In particular,
at the critical density $\alpha\beta=1$, the continuity of $Z_\alpha
g$ in Proposition \ref{prop:tpfft-properties} and Theorem
\ref{lem:zak-ineq}(a) show that $\cG(g,\alpha,\beta)$ has no positive
lower frame bound. Yet positive lower frame bounds may exist in the
discrete or the periodic setting, as the Zak transform $Z_\alpha g$ is
restricted to subsets of $[0,\alpha)\times[0,1/\alpha)$, as in Theorem
\ref{lem:zak-ineq} (b)--(d).  In Lemma \ref{lem:zeros-of-zak},
negative conclusions about the lower frame bound were given for the
case where $g$ is a continuous even function.  Here, we present
positive results about Gabor frames at the critical density if $g$ is
a TPFFT.  We start with the following example.

\begin{example}
The function $g(x)=\frac{1}{2}e^{-|x|}$ is  even and totally positive  of 
finite type $m=2$.
Its Fourier transform has the form 
\eqref{eq:tpfft-explicit-formula}, with
parameters  $\delta_{1,2}=\pm 1$ and $C_{1,2}=\frac{1}{2}$. 
Corollary \ref{cor:tpfft-zak}
gives the Zak transform
\[
    Z_\alpha g(x,\xi) = 
    \frac{e^{-x}}{2(1- e^{-\alpha (1 + 2 \pi j \xi)})} -
    \frac{e^{x }}{2(1- e^{\alpha (1- 2 \pi j \xi)})}.
\]
The only zero of $Z_ \alpha g$ which lies in the domain $[0,\alpha)\times[0,1/\alpha)$
is $(\frac{\alpha}{2},\frac{1}{2\alpha})$. Therefore, by  Theorem \ref{lem:zak-ineq}
we find that 
  \begin{itemize}
  \item \(\cG( Sg,\alpha,\beta)\) is  a Gabor frame for $\lt(\bZ)$ if $M$ is odd.
  \item \(\cG( P_Kg,\alpha,\beta)\) is  a Gabor frame for $\Lt(\bT_K)$ if $K/M$ is odd.
  \item \(\cG( P_KSg,\alpha,\beta)\) is  a Gabor frame for $\bC^K$ if $M$ is odd or  $K/M$ is odd.
  \end{itemize}
\end{example}

In a separate work it was shown  by T. Kloos and one of the authors 
\cite{KloosStoeckler13}
that
for every totally positive function $g$ of finite type $m\ge 2$ 
and $\alpha>0$, the Zak transform $Z_\alpha g$
has exactly one zero in the domain $[0,\alpha)\times[0,1/\alpha)$, and this
zero is located  at  $(x,\frac{1}{2\alpha})$ for some $x\in (0,\alpha)$. 
Based on this result, we can draw the following conclusion from Theorem \ref{lem:zak-ineq}.

\begin{theorem}\label{thm:criticalframe}
  Let \(g\) be a totally positive function of finite type $m\ge 2$. 
	 Assume $\alpha=M\in\bN$ and let $\beta=1/M$ and $K\in\bN$ such that $K/M\in\bN$. 
  \begin{itemize}
  \item If $K/M$ is odd, then  \(\cG( P_Kg,\alpha,\beta)\) is  a Gabor frame for $\Lt(\bT_K)$. 
  \item  If  $K/M$ is odd, then \(\cG( P_KSg,\alpha,\beta)\) is  a Gabor frame for $\bC^K$.
  \end{itemize}
	In addition, assume   that now  $g$ is even.  Then
  \begin{itemize}
  \item If $M$ is odd, then \(\cG( Sg,\alpha,\beta)\) is  a Gabor frame for $\Lt(\bT_K)$.
  \item  If  $M$ is odd, then \(\cG( P_KSg,\alpha,\beta)\) is  a Gabor frame for $\bC^K$.
  \end{itemize}
\end{theorem}

\begin{IEEEproof}
If $K/M$ is odd, then  $l/K\neq 1/(2M)$ for the  frequency parameter
in \eqref{eq:zak-ineq-TK}  for all
$ l \in \{0,1,\ldots,\frac{K}{M}-1\}$, hence the infimum in 
\eqref{eq:zak-ineq-TK} is positive. This shows that \(\cG( P_Kg,\alpha,\beta)\)
has a positive lower frame bound. The upper frame bound exists since $|Z_\alpha g|^2$ is
a continuous function and attains its maximum on the compact set $[0,\alpha]\times[0,1/\alpha]$.
The same argument works for \(\cG( P_KSg,\alpha,\beta)\).

If $g$ is even, then the unique zero of $Z_\alpha g$ inside the fundamental domain
$[0,\alpha)\times[0,1/\alpha)$ is at $(\frac{M}{2},\frac{1}{2M})$. 
If $M$ is odd,  then  $k\neq M/2$ for the time parameter
in \eqref{eq:zak-ineq-lt}  for all
$ k \in \{0,\ldots,M-1\}$. By continuity of $|Z_\alpha g|^2$, the infimum in 
\eqref{eq:zak-ineq-lt} is positive. This shows that \(\cG( Sg,\alpha,\beta)\)
is a Gabor frame for $\lt(\bZ)$. The same argument works for \(\cG( P_KSg,\alpha,\beta)\).
\end{IEEEproof}

\subsection{Algorithm for the computation of \(\gamma\)}
\label{sec:algo-gamma}
A special advantage of Gabor frames with TPFFT is the existence of a
dual window with compact support. We now describe an algorithm for the
computation of many possible dual windows with compact support.
This algorithm is an adaptation of the algorithm for the computation of
\(\gamma\) given in~\cite{grst13}. It is numerically much more stable than
the original one. The proof that this algorithm defines a dual window $\gamma$
for the Gabor frame $\cG(g,\alpha,\beta)$ is given in the appendix.

\begin{algorithm}
\label{algo:gamma}
Input parameters are the TPFFT $g$, the lattice parameters $\alpha ,\beta > 0$ with
$\alpha\beta<1$,
a parameter $L\in\bN_0$ controlling the support size of the dual window $\gamma$,
and a point $x\in[0,\alpha)$.
More precisely, $g$ is defined by specifying the vector 
$\mathbf{\delta}=(\delta_1,\ldots,\delta_{m+n})$ of its non-zero parameters
$\delta_k$ (see \eqref{eq:tpfft-fourier-transform}),
where $m,n\in\bN_0$, $m+n\ge 2$, and $m$ (resp. $n$) is the number of 
positive (resp. negative) parameters $\delta_k$. 

Output parameters are integers $i_1,i_2$ and the vector of values $\gamma(x+\alpha i)$,
$i_1\le i\le i_2$, in the support of $\gamma$.

  \begin{enumerate}
  \item Set $r:=\left\lfloor \frac{1}{1-\alpha \beta} \right\rfloor$.
  \item Set
    \[
      k_1 = -(r+1)m-L,\qquad
      k_2 = (r+1)n+L.
    \]
	 \item Set
    \[
      i_1 := \left\lfloor \frac{k_1 + m - 1}{\alpha\beta}-\frac{x}{\alpha}\right\rfloor+1,\qquad
      i_2 := \left\lceil \frac{k_2 - n + 1}{\alpha\beta}-\frac{x}{\alpha}\right\rceil-1.
    \]
   \item Set
    \begin{equation*}
      P = (p_{ik})_{i_1\le i \le i_2,~ k_1\le k \le k_2}\quad \hbox{with}
			\quad  p_{i,k}= g \left(x+\alpha i- \frac{k}{\beta}\right).
    \end{equation*}
  \item Compute the  
	pseudoinverse $P^\dagger=(q_{ki})_{k_1\le k \le k_2,~ i_1\le i \le i_2}$ of $P$.
  \item Take the row with index \(k=0\) of $P^\dagger$. Its coefficients 
    define the values of the dual window $\gamma$ at the points $\{x+\alpha i \,|\,i_1 \le
    i \le i_2\}$, i.e.
    \begin{equation}\label{eq:set-gamma}
    \gamma(x+\alpha i):=\begin{cases} \beta q_{0,i},&\text{if } i_1\le i \le i_2,\\
		0&\text{if } i<i_1~\text{or } i>i_2.
		\end{cases}
    \end{equation}
  \end{enumerate}
\end{algorithm}
This algorithm yields a class of dual windows \(\gamma\) (depending on $L$)
 for the Gabor frame \(\cG(g,\alpha,\beta)\) in
\(\Lt(\bR)\). Note that the function $\gamma$ is piecewise continuous.
Indeed, the only discontinuity (with respect to $x$) can occur when the 
selection of $i_1$ or $i_2$ in step 3 of the algorithm 
has a jump, since all entries of the matrix
$P$ and its pseudoinverse $P^\dagger$ depend continuously on $x$. 
Moreover, the definitions in the algorithm imply that
\[
   x+\alpha i_1\ > x+\alpha\left( \frac{k_1 + m - 1}{\alpha\beta}-
	\frac{x}{\alpha}\right) = \frac{-rm-L- 1}{\beta},
\]
with $r:=\left\lfloor \frac{1}{1-\alpha \beta} \right\rfloor$, and
likewise  $ x+\alpha i_2 < \frac{rn+L+ 1}{\beta}$.
Therefore, the input parameter $L$ is used to control the 
	support of $\gamma$, 
        \begin{equation}
          \label{eq:ch23}
           {\rm supp}\,\gamma \subset \left[\frac{-rm-L- 1}{\beta},
		\frac{rn+L+ 1}{\beta}\right].
        \end{equation}

When working with discrete, periodic or finite signals,
we need to sample or periodize \(\gamma\), or apply both
operations. For example, in order to obtain a discrete dual window
for the Gabor frame $\cG(Sg,a,\frac{1}{M})$ of $\lt(\bZ)$, where $a,M\in\bN$,
we choose \(x \in \{0,1,\ldots,a-1\}\) in Step 3 of Algorithm \ref{algo:gamma}
and obtain the dual window $S\gamma$. 
Additional periodization with period of length $K$ provides the dual window
 \(P_K S\gamma\) of the Gabor frame $\cG(P_KSg,a,\frac{1}{M})$ of $\bC^K$,
provided that $K/a\in\bN$ and $K/M\in\bN$.

\section{Examples}
\label{sec:examples}

\begin{figure}[!t]
\centering
\subfloat[][]{\includegraphics[width=2.2in]{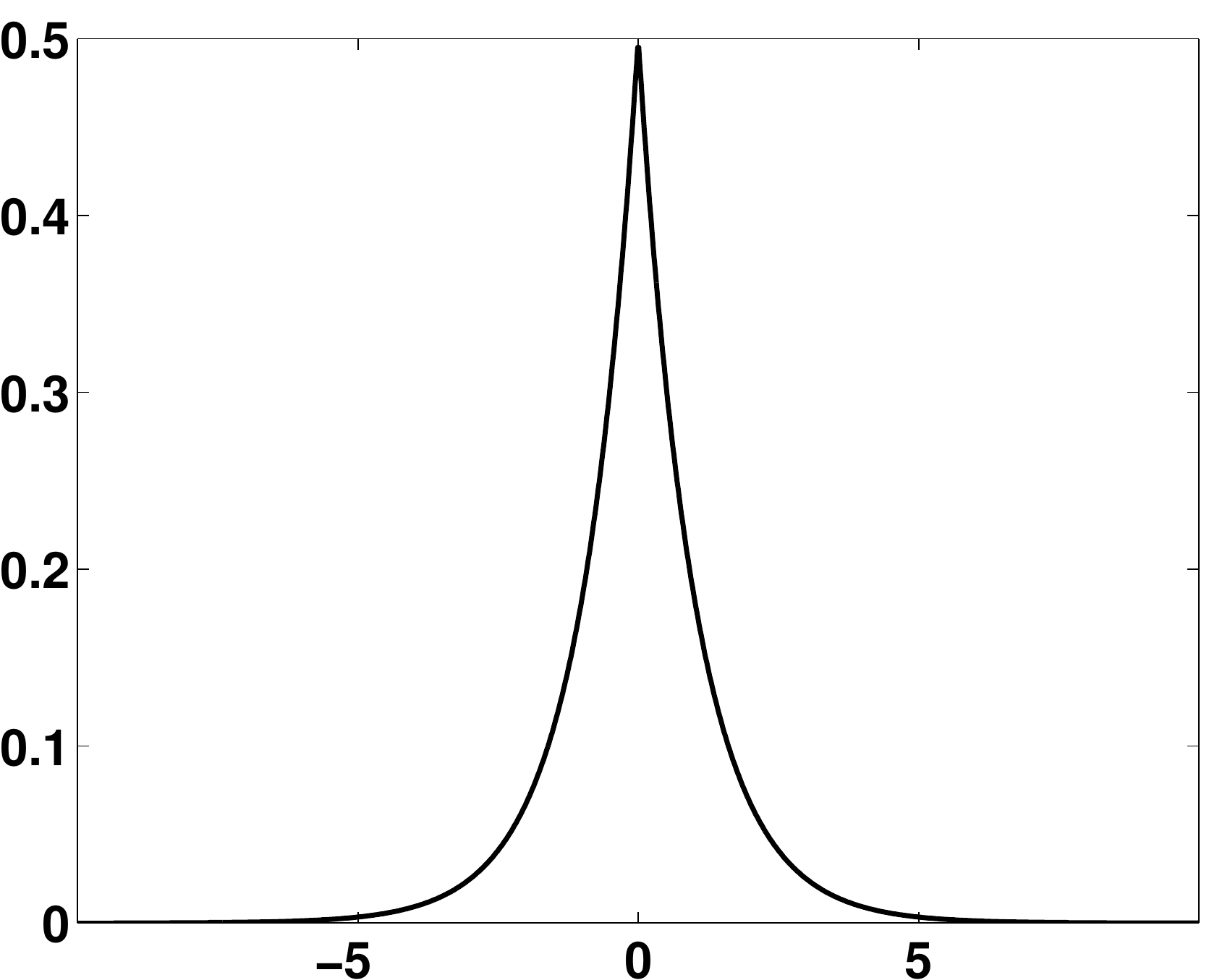}%
\label{fig:two_sided_exp}}
\quad
\subfloat[][]{\includegraphics[width=2.2in]{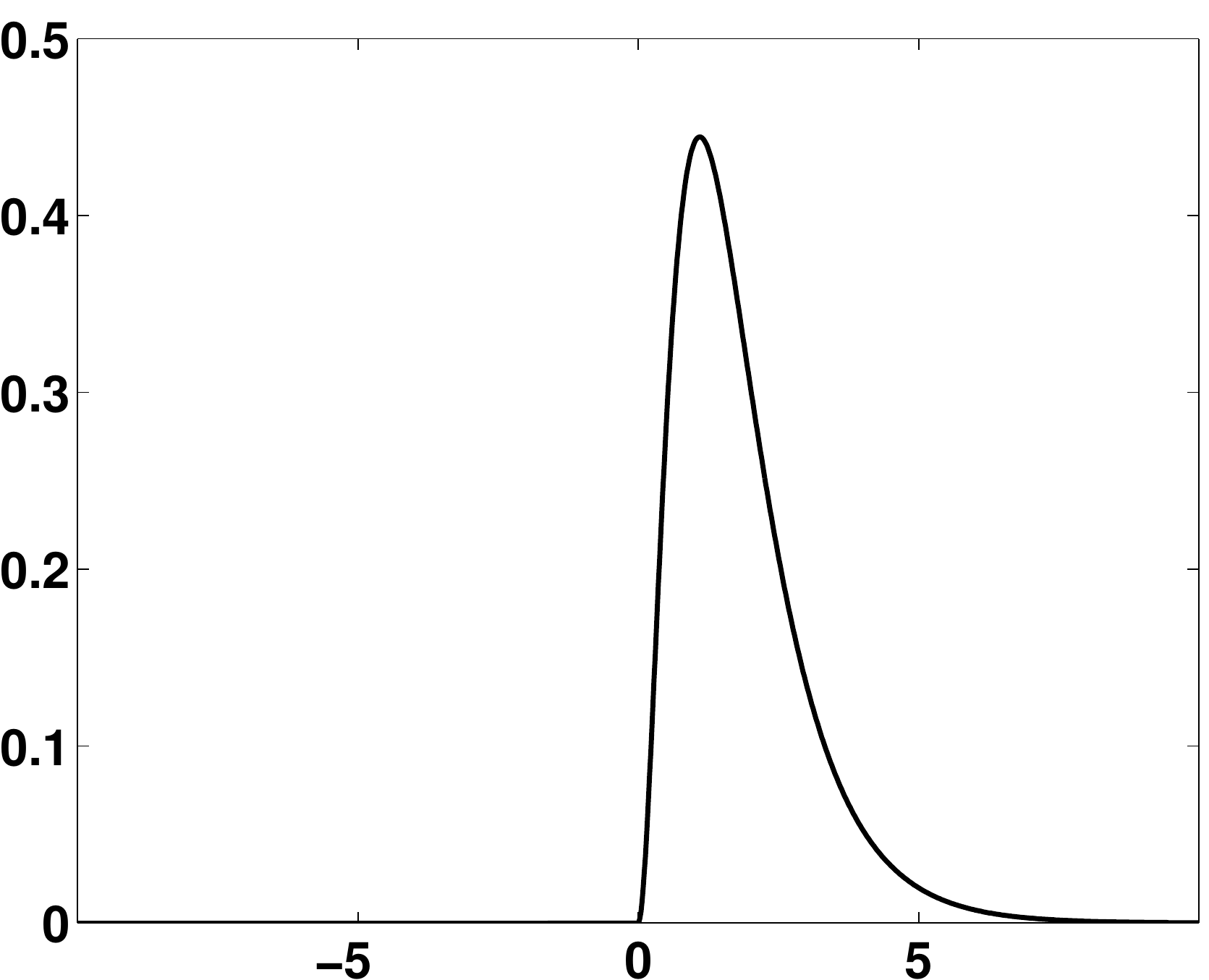}%
\label{fig:onesided}}\\
\subfloat[][]{\includegraphics[width=2.2in]{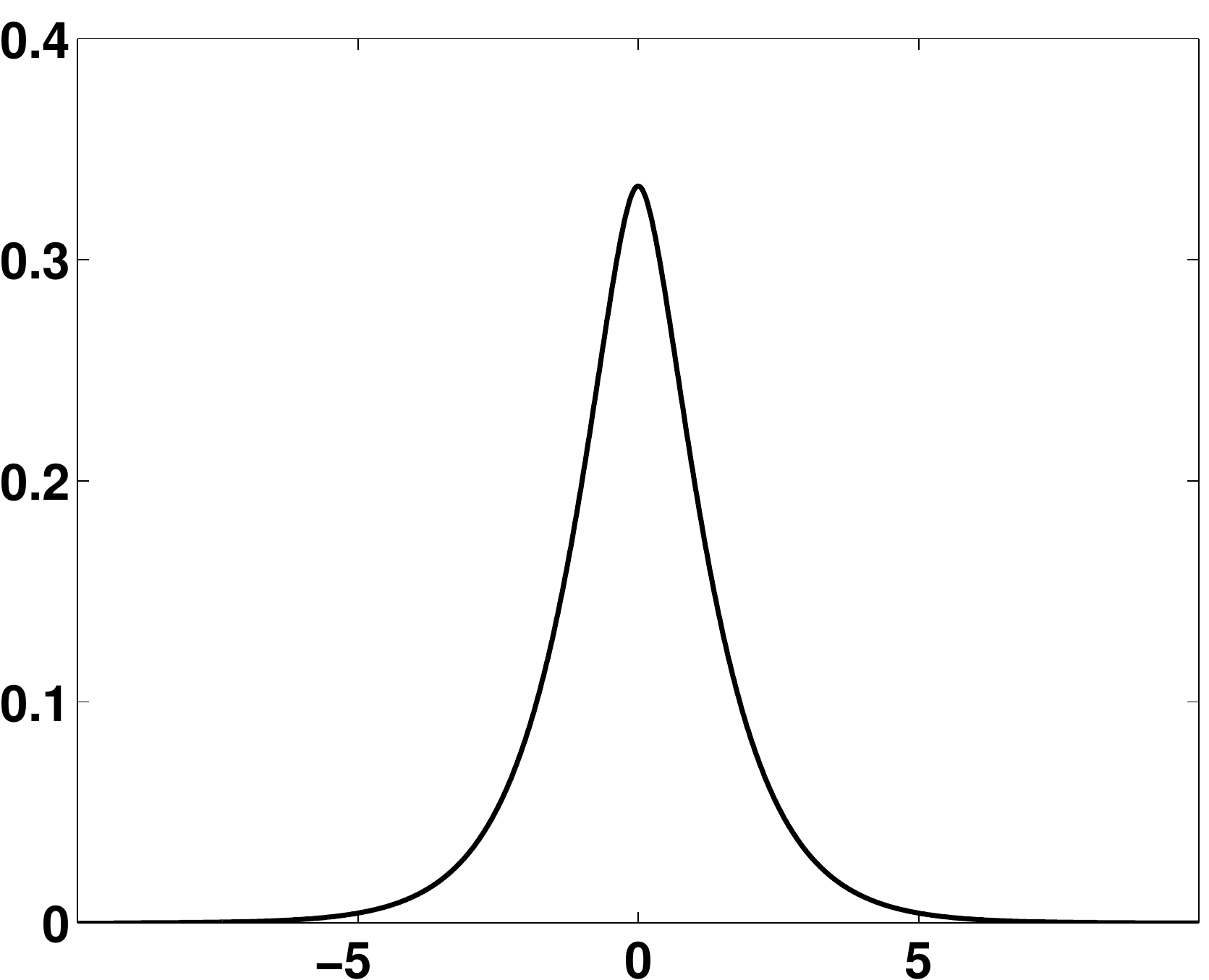}%
\label{fig:g_stoeckler}}
\quad
\subfloat[][]{\includegraphics[width=2.2in]{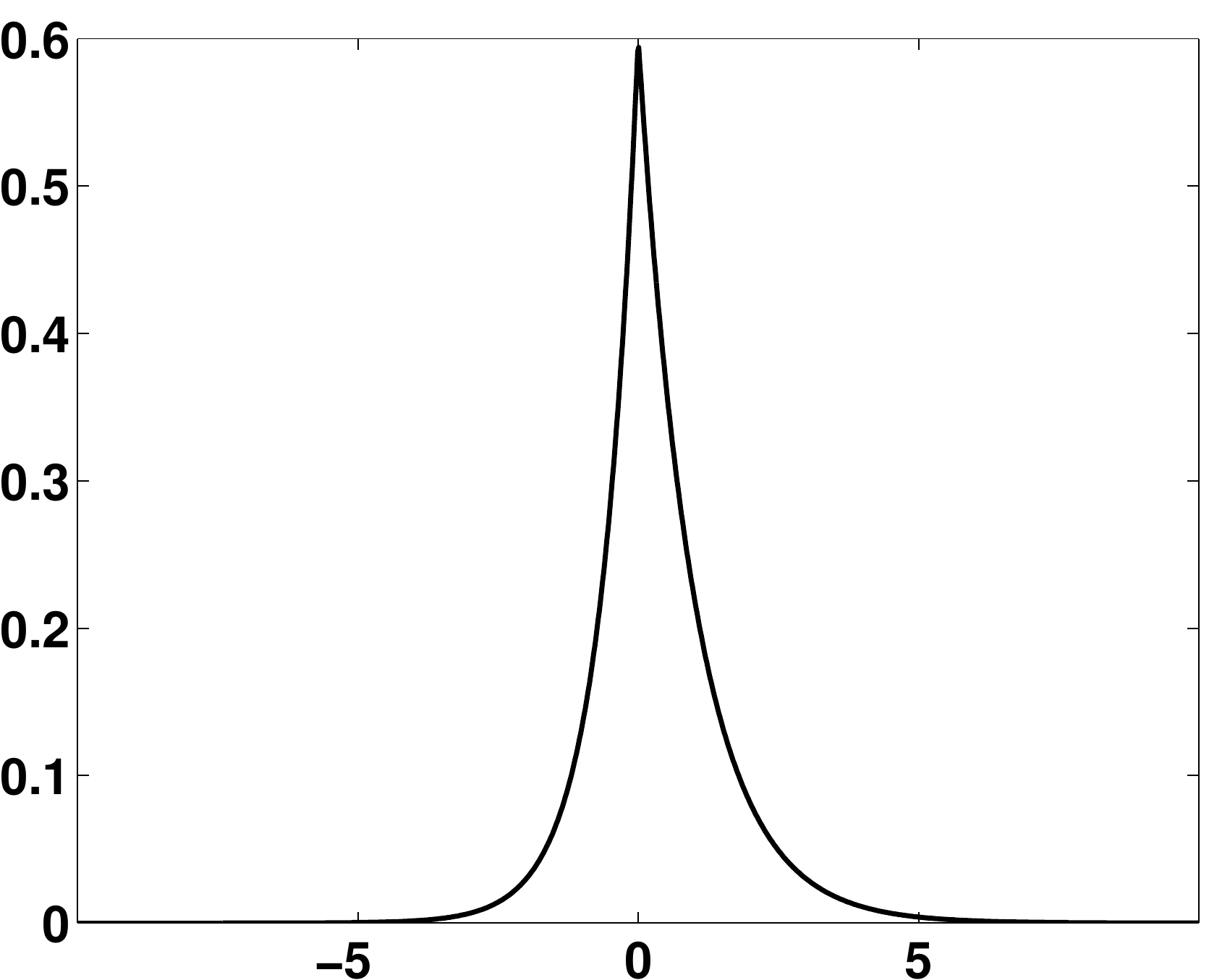}%
\label{fig:asymmetric_exp}}
\caption{\small Four examples of totally positive functions of finite
  type. 
	(a)
  \(g_1(x)=\frac{1 }{2}e^{-|x|}\); 
	(b) \(g_2(x)=(3e^{-3x} - 6 e^{-2 x}+3 e^{-x})h(x)\); 
	(c) \(g_3(x) = \frac{2}{3} e^{-|x|}-\frac{1}{3}e^{-2 |x|}\); 
	(d) \( g_4(x) =e^{3x}h (-x) + e^{-2x} h(x)\).}
\label{fig:tpfft_examples}
\end{figure}

Fig. \ref{fig:tpfft_examples} shows four totally positive functions of
finite type. Fig. \ref{fig:tpfft_examples}\subref{fig:two_sided_exp}
shows the twosided exponential function \(g_1(x) = \frac{1}{2}\,e^{-|x|}\) with
parameters \(\delta_1 = -1,\, \delta_2 = 1\). This is an even  TPFFT of type $2$,
and $m=1$, $n=1$ in Algorithm \ref{algo:gamma}. 
Fig. \ref{fig:tpfft_examples}\subref{fig:onesided} depicts
the function \(g_2(x)=(3 e^{-3x} - 6 e^{-2 x}+3
e^{-x})h(x) \) with parameters \(\delta_1 = 1,\, \delta_2 =
1/2,\, \delta_3 =1/3\) and type
\(m=3\); here $n=0$ as all $\delta_k$ are positive, and ${\rm supp}\, g_2=[0,\infty)$. 
Fig. \ref{fig:tpfft_examples}\subref{fig:g_stoeckler} shows
the even function \(g(x) = \frac{2}{3} e^{-|x|}-\frac{1}{3} e^{-2 |x|}\) of type $4$ (with \(m=2,\,n=2\))
with parameters \(\delta_{1,2} = \pm 1,\, \delta_{3,4} = \pm 1/2\). Finally,
Fig. \ref{fig:tpfft_examples}\subref{fig:asymmetric_exp} depicts the
asymmetric exponential \( g_4(x) = e^{3x}h (-x) + e^{-2x}
h(x)\) with parameters \(\delta_1 = -2/3\) and \( \delta_2 =
1\), here \(m=1,\,n=1\).

\begin{figure}[!t]
\centering

\subfloat[][]{\includegraphics[width=2.2in]{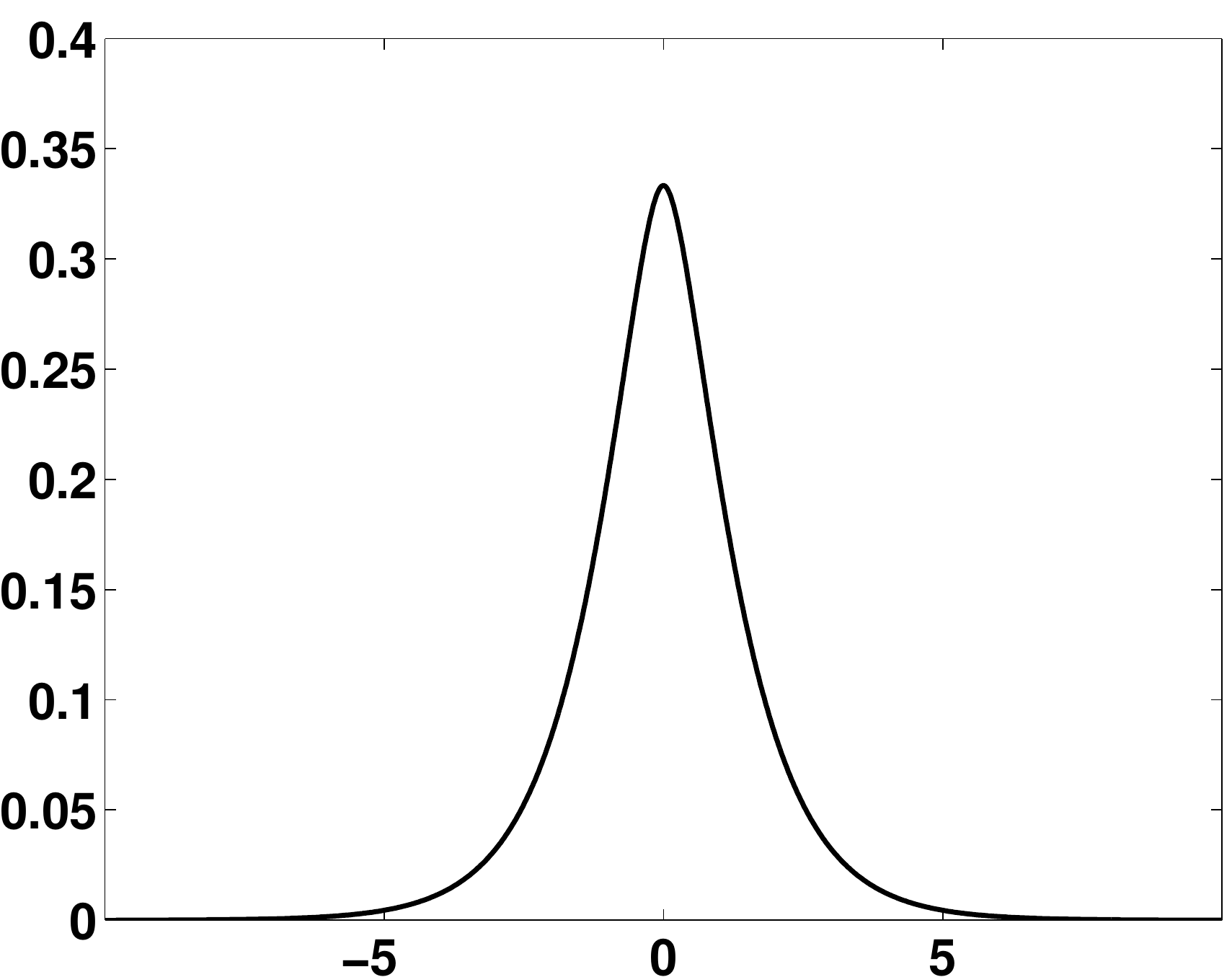}%
\label{fig:g_stoeckler_2}}
\quad
\subfloat[][]{\includegraphics[width=2.2in]{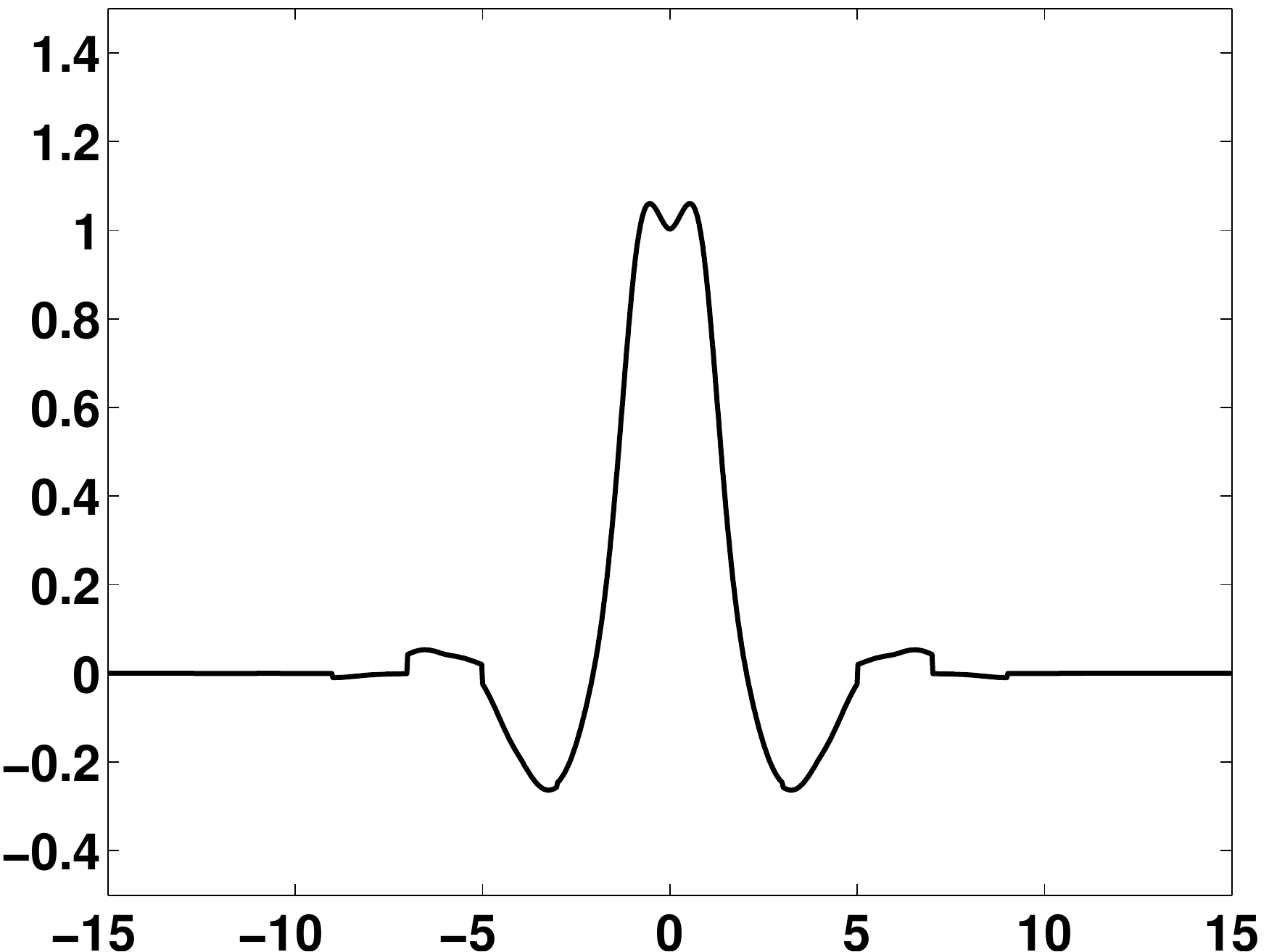}%
\label{fig:stoeckler_dual_1}}\\

\subfloat[][]{\includegraphics[width=2.2in]{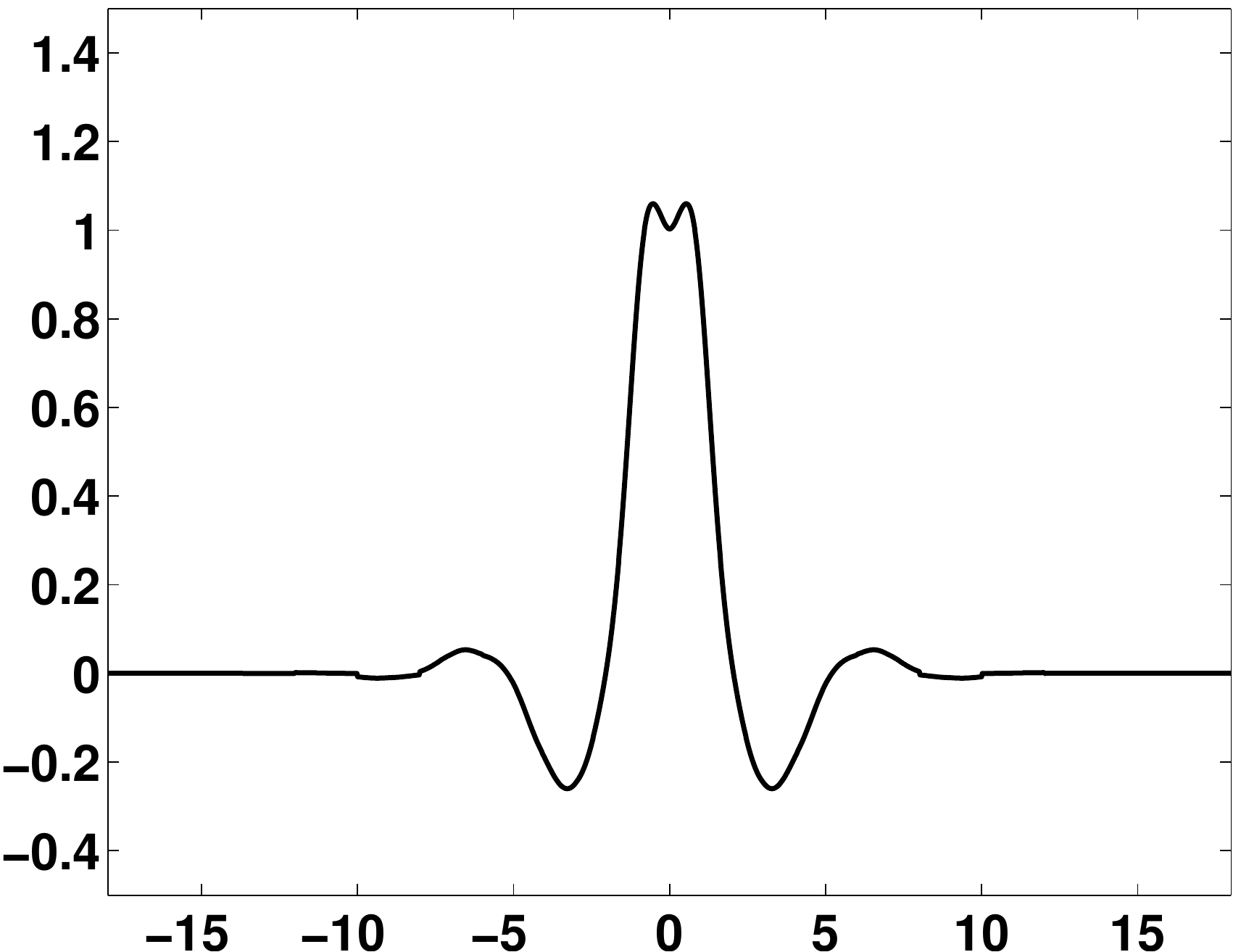}%
\label{fig:stoeckler_dual_2}}
\quad
\subfloat[][]{\includegraphics[width=2.2in]{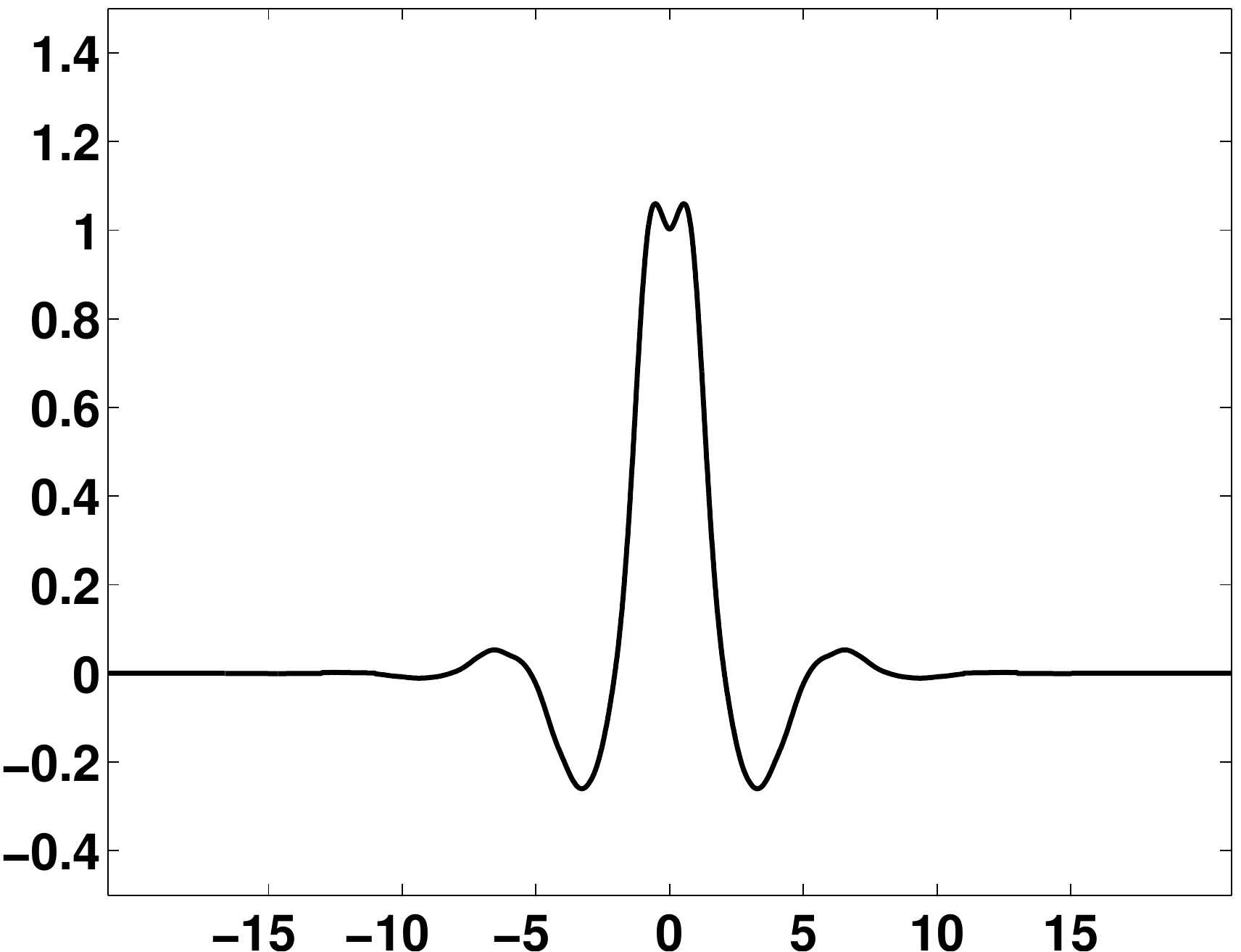}%
\label{fig:stoeckler_dual_3}}
\caption{\small The function \(g(x) = \frac{2}{3} e^{-|x|}-\frac{1}{3} e^{-2 |x|}\)
  and three dual windows obtained by successively increasing \(L\)
  in Algorithm \ref{algo:gamma}.}
\label{fig:tpfft_duals}
\end{figure}

Fig. \ref{fig:tpfft_duals} shows the function \(g(x) = \frac{2}{3}
e^{-|x|}-\frac{1}{3} e^{-2 |x|}\) and three of its dual windows. The lattice
parameters are
\(\alpha=2,\,\beta=1/3\). Fig. \ref{fig:tpfft_duals}\subref{fig:stoeckler_dual_1}
is the dual window as computed by Algorithm \ref{algo:gamma} with
\(L=0\) giving the smallest support, but obvious discontinuities of $\gamma$. 
Fig. \ref{fig:tpfft_duals}\subref{fig:stoeckler_dual_2} and
\ref{fig:tpfft_duals}\subref{fig:stoeckler_dual_3} are computed for $L=1$ and $L=2$.

\begin{figure}[!t]
\centering

\subfloat[][]{\includegraphics[width=2.2in]{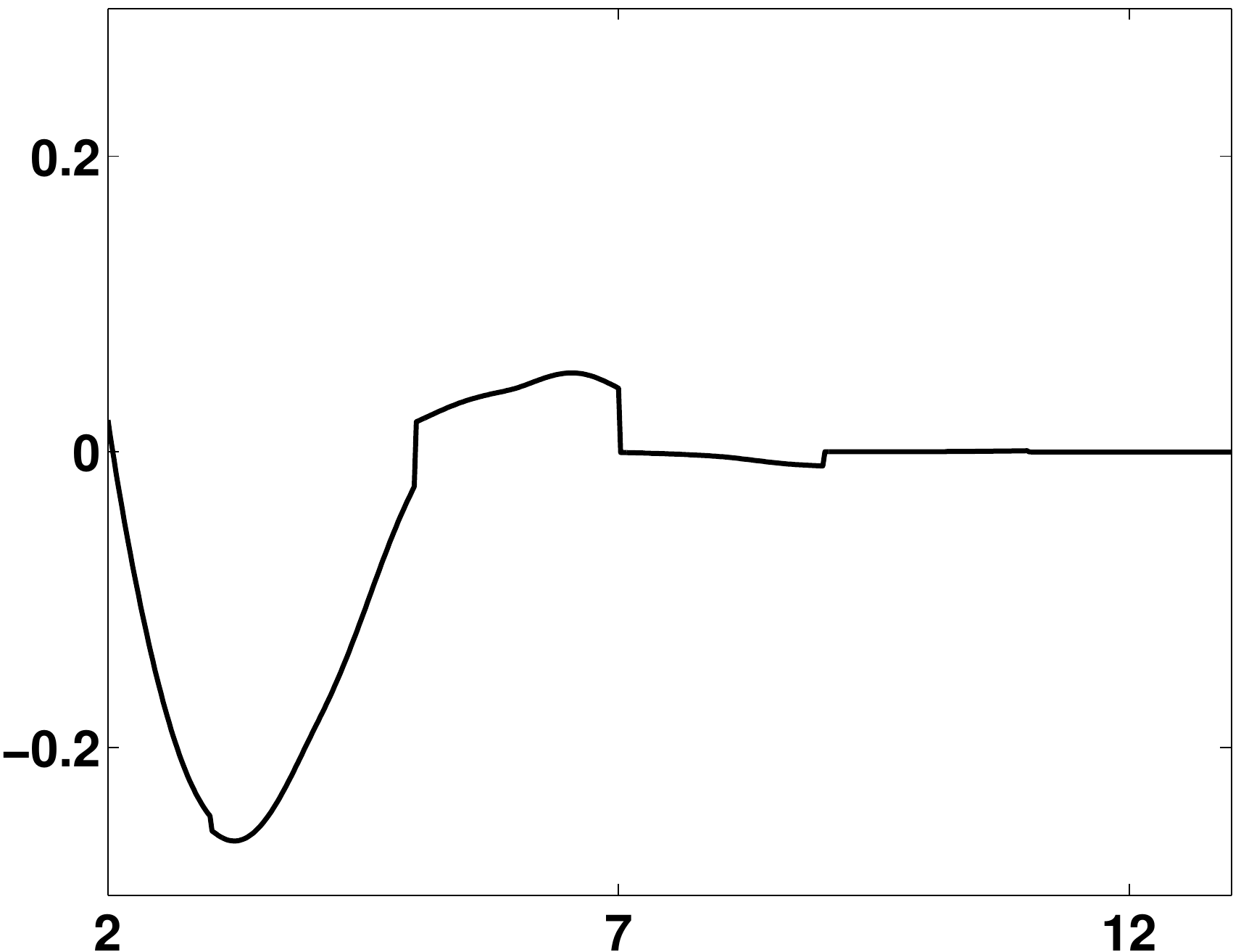}%
\label{fig:closeup_dual_L0}}
\quad
\subfloat[][]{\includegraphics[width=2.2in]{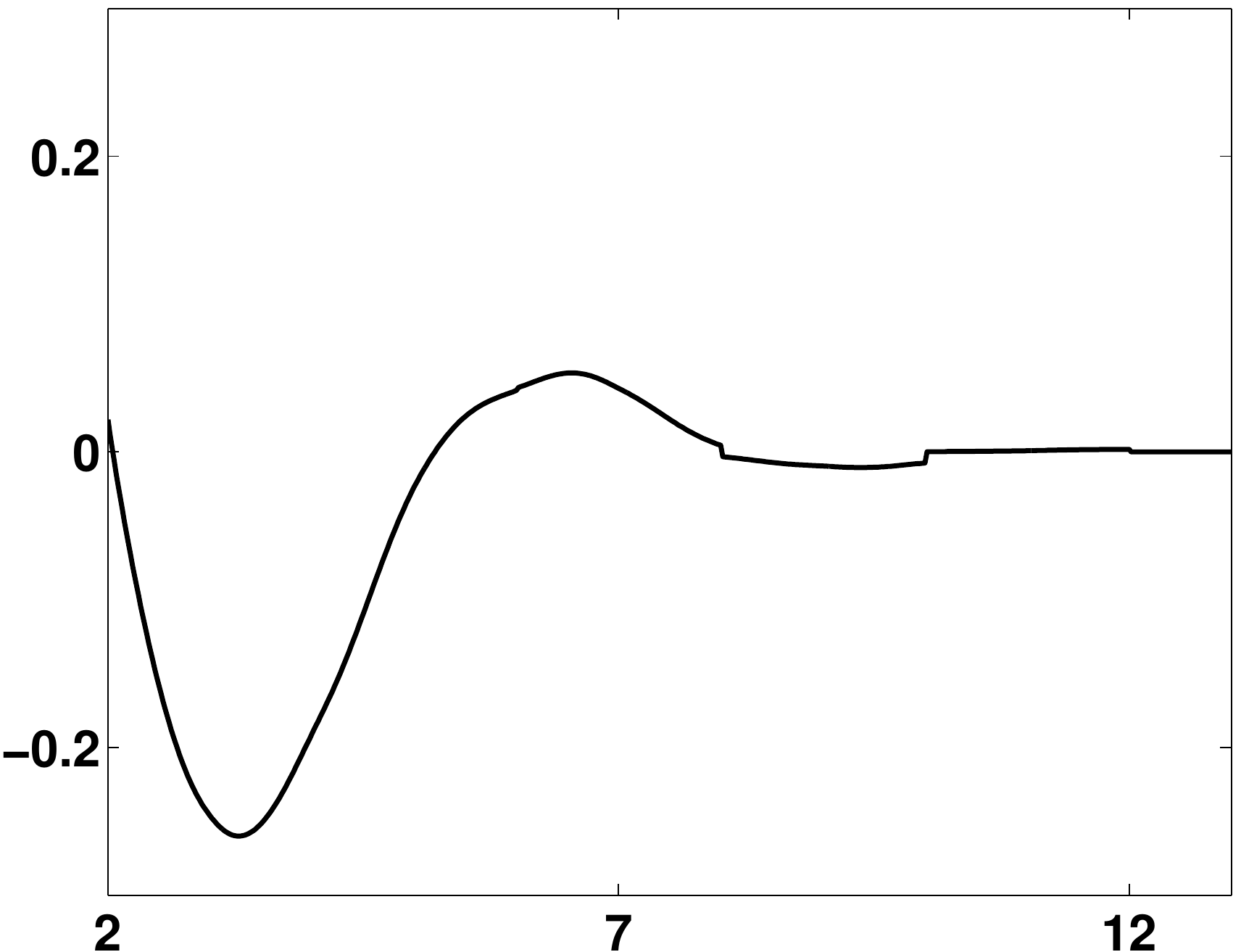}%
\label{fig:closeup_dual_L1}}\\

\subfloat[][]{\includegraphics[width=2.2in]{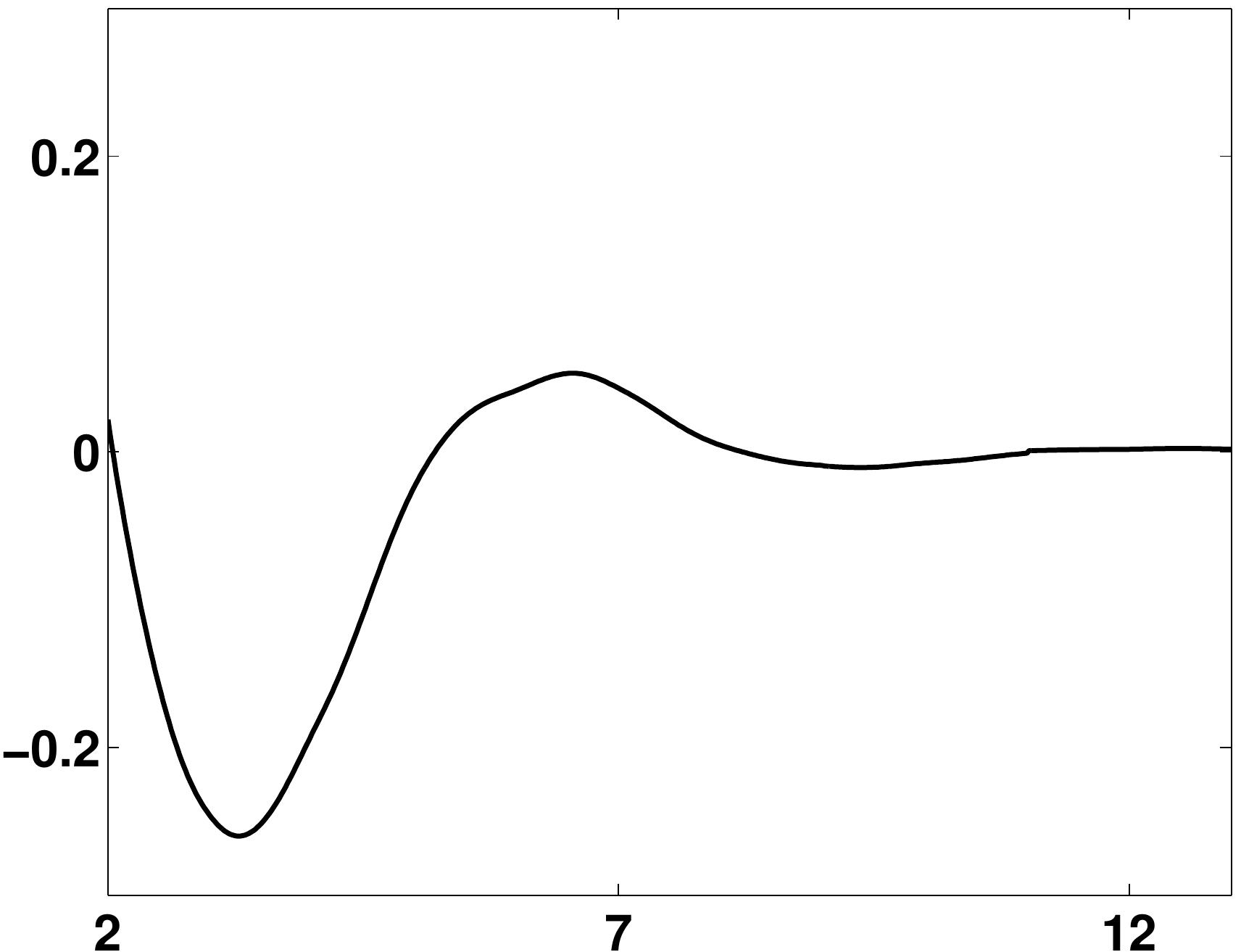}%
\label{fig:closeup_dual_L2}}
\quad
\caption{\small A closeup of the discontinuities of the dual windows
  in Figure \ref{fig:tpfft_duals}.}
\label{fig:tpfft_duals_closeup}
\end{figure}

The free parameter $L$ in the algorithm not only determines the length
of the support of the dual window as in~\eqref{eq:ch23}, it also seems
to parametrize the smoothness of the dual window, see Fig.
\ref{fig:tpfft_duals_closeup}. With increasing $L$ the size of the
jumps decreases. As $L$ tends to infinity, the dual window converges
to the canonical dual window. At this time we do not have a rigorous
proof to confirm these numerical observations.

\section{Conclusion}
We studied totally positive functions as a class of windows for Gabor
frames (Weyl-Heisenberg frames). Such windows are universal in two
ways: 

(a) Oversampling the time-frequency plane, even by a minimal amount,
guarantees the frame property. If $\alpha \beta <1$, the Gabor family
$\mathcal{G}(g,\alpha , \beta )$ is a frame (without any further
restrictions).  Likewise, undersampling the time-frequency plane, even
by a minimal amount, guarantees the linear independence. If $\alpha
\beta >1$, then the Gabor family $\mathcal{G}(g,\alpha , \beta )$ is a
Riesz sequence.  The case of the critical density is also considered.

(b) Totally positive functions can be used as windows for Gabor frames
in the setting for time-continuous signals, for time-discrete signals,
for continuous periodic signals, and for discrete periodic signals.

In addition, a totally positive window decays exponentially and thus
its numerical support is finite. We give a simple numerical algorithm
for the exact computation of dual windows with compact support.


%

\appendices
\section{}
For the proof that Algorithm \ref{algo:gamma} computes the values of a
dual window $\gamma$ we make use of the following results,
see~\cite{grst13,janssen95,ron-shen97} and the references cited.
\begin{theorem}[Characterization of Gabor frames]
  \label{thm:char-gabor-frames}
  Let $g\in L^2(\mathbb{R})$ and $\alpha,\beta > 0$. Then the
  following are equivalent.
  \begin{enumerate}
  \item $\mathcal{G}(g,\alpha,\beta)$ is a frame for
    $L^2(\mathbb{R})$.
  \item There exists a so-called dual window $\gamma$, such that
    $\mathcal{G}(\gamma,\alpha,\beta)$ is a Bessel sequence and
    satisfies the \emph{Wexler-Raz biorthogonality relations}
    \begin{equation}
      \label{eq:wexler-raz}
      \langle\gamma , M_{l/\alpha}T_{k/\beta} g\rangle = \alpha \beta
      \delta_{k,0} \delta_{l,0}, \quad \forall k,l \in \mathbb{Z}.
    \end{equation}
  \end{enumerate}
\end{theorem}
Analogous characterizations hold for Gabor frames on $\ell ^2(\bZ )$,
on $L^2(\bT _K)$ and $\bC ^L$. 

\begin{theorem}[The Schoenberg-Whitney conditions~\cite{SW53}]
  Let \(g\in\Lo(\bR)\) be a TPFFT.  In the factorization
  (\ref{eq:finite-factorization}) of \(\hat{g}\)  let  $m\in \bN$  denote the
  number of positive  \(\delta_\nu\)  and   $n\in \bN$  denote the
  number of negative  \(\delta_\nu\) respectively. For
  two sequences \((x_i)_{i=1,\ldots,N}\) and
  \((y_k)_{k=1,\ldots,N},\,N \in\bN\), the determinant
  \(\det(g(x_i-y_k)_{i,k=1, \ldots,N})\) is strictly positive, if and
  only if
  \begin{equation}
    \label{eq:schoenberg-whitney-conditions}
    y_{i-n} < x_i < y_{i+m} \quad \mbox{ for }1\le i\le N
  \end{equation}
	(with the interpretation $y_k=-\infty$ for $k\le 0$ and 
	$y_k=\infty$ for $k>N$.)
\end{theorem}

Now we prove that Algorithm \ref{algo:gamma} computes the function $\gamma$
which satisfies the Wexler-Raz biorthogonality relation.
\begin{IEEEproof}
Recall that $\alpha\beta<1$
	and \(r=\lfloor
  \frac{1}{1-\alpha\beta} \rfloor\), which implies
  $  r+1 >  (1-\alpha\beta)\inv$.
We consider the matrix $P$ 
in Step 4 of Algorithm \ref{algo:gamma},
    \begin{equation}
      \label{eq:Pxinproof}
      P:= \left(g\big(x+\alpha i - 
			\tfrac{k}{\beta}\big)\right)_{i_1\le i\le i_2,~ k_1\le k\le k_2}.
    \end{equation}
The number $M:=i_2-i_1+1$ of rows of $P$ 
exceeds the number $N:=k_2-k_1+1$ of columns, as by simple computations
\begin{align*}
   i_2-i_1& \geq \left(\frac{k_2-n+1}{\alpha\beta}-\frac{x}{\alpha}-1\right)
	-\left(\frac{k_1+m-1}{\alpha\beta}-\frac{x}{\alpha}+1\right)\\
	&=\frac{k_2-k_1}{\alpha\beta}-\frac{n+m-2}{\alpha\beta}-2\\
	&> (k_2-k_1)+\frac{1}{\alpha\beta}\left((k_2-k_1)(1-\alpha\beta) -(n+m)\right)
\end{align*}
and 
\[
   (k_2-k_1)(1-\alpha\beta)  \ge (n+m)(r+1)(1-\alpha\beta)>n+m.
\]
\begin{claim}\label{claim:one}  $P$ has full rank $N$. 
\end{claim}
\begin{IEEEproof}
Numbering rows of $P$ 
	from $1$ to $M$
	and columns from $1$ to $N$, as usual, the matrix entries are
	$p_{ik}=g(x_i-y_k)$, where
\begin{equation}\label{eq:defxi}
   x_i:= x+(i_1-1+i)\alpha,\qquad y_k:=\frac{k_1-1+k}{\beta},
\end{equation}
for $i=1,\ldots,M$ and $k=1,\ldots,N$. 
Since  $\alpha\beta<1$, 
each interval $(y_k,y_{k+1})$ contains at least one $x_i$.
The definition of $i_1,i_2$ gives 
\begin{equation}\label{eq:SW-endpoint}
    y_{m}<x_1<y_{m+1},\qquad y_{N-n}<x_M<y_{N-n+1}
\end{equation}
(with  
definitions of $y_0,y_{N+1}$ as in \eqref{eq:defxi} if $m=0$ or $n=0$). 
As elaborated in the proof
  of Theorem 8 in~\cite{grst13}, we can select a subsequence $(x_{i_l})_{1\le l\le N}$ of $(x_i)$,
\[
   x_1=:x_{i_1} <x_{i_2}<\cdots  <x_{i_N}:=x_M
\]
such that the Schoenberg-Whitney conditions \eqref{eq:schoenberg-whitney-conditions}
are satisfied for this subsequence 
intertwined with the sequence $y_1,\ldots,y_N$. Then the
corresponding $N\times N$ submatrix  consisting of the  rows $i_1,\ldots,i_N$ of $P$
 has nonzero determinant, and thus $P$ has full rank.
\end{IEEEproof}

Claim \ref{claim:one} shows that the
pseudo-inverse $P^\dagger$ is a left inverse of $P$ 
satisfying $P^\dagger P=I$. We let $k_0:=1-k_1$ be the ``central'' 
column index
in $P$ associated with $y_{k_0}=0$. The corresponding row of $P^\dagger$ is
denoted by 
\begin{equation}\label{eq:gamma-row}
     v=(q_{k_0,i})_{1\le i\le M}.
\end{equation}
Furthermore, define the vector $w_k\in \bR ^m$ by  
\begin{equation}\label{eq:wk}
   w_k:=\left(g\big(x_i-\tfrac{k}{\beta}\big)\right)_{1\le i\le M},\qquad k\in\bZ.
\end{equation}
\begin{claim}\label{claim:two}
   Every column vector  
	$w_k$ 
        with $k<k_1$ 	lies in the linear span of the first $m$ columns of
        $P$. Likewise, $w_k$ with  $k>k_2$
	lies in the linear span of the last  $n$ columns of $P$.
\end{claim}
\begin{IEEEproof} If $m=0$ and $k<k_1$, the claim means that $w_k=0$. 
Indeed, for $m=0$, the support of $g$ is $(-\infty,0]$ by
\eqref{eq:tpfft-explicit-formula}, 
and we have 
$ x_1=x+\alpha i_1>\frac{k_1-1}{\beta}$.
Since $(x_i)$ is increasing, 
we obtain $g\left(x_i-\frac{k}{\beta}\right)=0$ for all $1\le i\le M$. 

If $m\ge 1$ and $k<k_1$, we define the $M\times (m+1)$ matrix
\[
   P_0= \left(g\big(x_i-\tfrac{\tilde k}{\beta}\big)\right)_{1\le i\le M,~
	\tilde k=k,k_1,\ldots,k_1+m-1}.
\]
The last $m$ columns of $P_0$ are the first $m$ columns of the
matrix $P$,  therefore they are linearly independent.
In order to prove the claim, we show that $\text{rank}(P_0)= m$.
Suppose $P_0$ has full rank $m+1$, then there exists an invertible
    $(m+1)\times (m+1)$ submatrix $P_1$ of $P_0$ with row
    indices $1\le s_1<s_2<\cdots <s_{m+1 }\le M$. 
That is,
$ P_1=\left(g(x_{s_i}-\eta_k)\right)_{1\le i,k\le m+1}$
where 
\[  \eta_1=\frac{k}{\beta}< \eta_2=y_1 <\cdots <\eta_{m+1}=y_m.
\]
Since $P_1$ is
		invertible, the Schoenberg-Whitney conditions imply
                that $x_1 \leq x_{s_1}<\eta_{m+1}=y_m$,
which contradicts \eqref{eq:SW-endpoint}.  The case $k>k_2$ is analogous. 
\end{IEEEproof}
	
Now we can complete the proof. 
The vector 
	$w_{0}$ is the ``central'' column of $P$ with column index
	$k_0=1-k_1$. 
The identity $P^\dagger P=I$ implies that
the row vector $v$ in \eqref{eq:gamma-row} satisfies $v\cdot w_{0}=1$ 
and $v\cdot w_k=0$ for $k_1\le k\le k_2$, $k\ne 0$. Now our  choice of
$k_1,k_2$  implies 
	$m<k_0<N-n+1$  and thus the orthogonality of the vector $v$
	to the first $m$ and the last $n$ columns of $P$. By Claim \ref{claim:two}
	the additional orthogonality relations $v\cdot w_k=0$ are
        satisfied  for all $k<k_1$,
	and all  $k>k_2$.
Recall that the definitions of $P$, $P^\dagger$, $v$ and $w_k$
depend on $x$. 
The definition of the function values of $\gamma$ in \eqref{eq:set-gamma}
can be written as
\[  
    \gamma(x_i):=\begin{cases} \beta q_{k_0,i},&\text{if } 1\le i \le M,\\
		0&\text{otherwise,} 
		\end{cases}
\] 
where $x_i=x+(i_1-1+i)\alpha$ for $i\in\bZ$ as in \eqref{eq:defxi}.
It was proved in \cite{grst13} that, by letting $x\in [0,\alpha)$, this 
defines a measurable function $\gamma\in \Lt(\bR)$, and that the Gabor family
$\cG(\gamma,\alpha,\beta)$ is a Bessel family in $\Lt(\bR)$. 
 It remains to check the Wexler-Raz biorthogonality relation. The definition of
$\gamma $ leads to
  \begin{align*}
    \langle \gamma, M_{l/\alpha}T_{k/\beta} g\rangle &=
    \int_{\bR}
    \gamma (x) \overline{g}\big(x-\tfrac{k}{\beta}\big) e^{-2 \pi jl x/\alpha}\,dx\\
    &=     \int_{0}^{\alpha} \sum_{i\in\mathbb{Z}}
    \gamma (x+i\alpha) \overline{g}\big(x+i\alpha-\tfrac{k}{\beta}\big) e^{-2 \pi jl x/\alpha}\,dx\\
    &= \beta \int_{0}^{\alpha} \delta_{k,0} e^{-2 \pi jl
      x/\alpha}\,dx = \alpha\beta \delta_{k,0} \delta_{l,0}.
  \end{align*}
	The sum in the second integral is finite; it represents the product $v(x)\cdot w_k(x)$ with 
	$v(x)$ in \eqref{eq:gamma-row} and $w_k(x)$ in \eqref{eq:wk} where the dependency on $x$ 
	is now included  in the notation. 
 Thus $\gamma$ is a dual window.
\end{IEEEproof}




\ifCLASSOPTIONcaptionsoff
  \newpage
\fi



\bibliographystyle{IEEEtran}

\def\cprime{$'$} \def\cprime{$'$} \def\cprime{$'$} \def\cprime{$'$}
  \def\cprime{$'$}

\end{document}